\documentclass[fleqn,usenatbib]{mnras}

\usepackage{newtxtext,newtxmath}
\usepackage[flushleft]{threeparttable}
\usepackage{lscape}

\usepackage{float,xcolor}
\usepackage{rotating}
\usepackage{tablefootnote}

\usepackage[T1]{fontenc}

\DeclareRobustCommand{\VAN}[3]{#2}
\let\VANthebibliography\thebibliography
\def\thebibliography{\DeclareRobustCommand{\VAN}[3]{##3}\VANthebibliography}

\usepackage{graphicx}	
\usepackage{amsmath}	
\usepackage{graphicx}
\usepackage{float}

\title[Polarimetric imaging of IRAS\,08544-4431.]{Multi-wavelength high-resolution polarimetric imaging of second-generation disc around post-AGB binary IRAS\,08544-4431 with SPHERE.}

\author[K. Andrych et al.]{
Kateryna Andrych\thanks{E-mail: kateryna.andrych@students.mq.edu.au}$^{1,2}$,
Devika Kamath$^{1,2,3}$,
Hans Van Winckel$^{4}$, 
Jacques Kluska$^{4}$,\newauthor
Hans Martin Schmid$^{5}$,
Akke Corporaal$^{6}$,
Julien Milli$^{7}$
\\
$^{1}$School of Mathematical and Physical Sciences, Macquarie University, Balaclava Road, Sydney, NSW 2109, Australia\\
$^{2}$Research Centre for Astronomy, Astrophysics and Astrophotonics, Macquarie University, Balaclava Road, Sydney, NSW 2109, Australia\\
$^{3}$INAF, Osservatorio Astronomico di Roma, Via Frascati 33, 00077, Monte Porzio Catone, Italy \\
$^{4}$Institute of Astronomy, KU Leuven, Celestijnenlaan 200D, 3001 Leuven, Belgium\\
$^{5}$Institute for Particle Physics and Astrophysics, ETH Zurich, Wolfgang-Pauli-Strasse 27, 8093, Zurich, Switzerland\\
$^{6}$European Southern Observatory, Alonso de Córdova 3107, Vitacura, Santiago, Chile\\
$^{7}$Université Grenoble-Alpes, CNRS, IPAG, 38000, Grenoble, France\\
}

\date{Accepted XXX. Received YYY; in original form ZZZ}

\pubyear{2024}

\begin{document}
\label{firstpage}
\pagerange{\pageref{firstpage}--\pageref{lastpage}}
\maketitle

\begin{abstract}

Exploring the formation and evolution of second-generation circumbinary discs around evolved binary stars, such as post-Asymptotic Giant Branch (post-AGB) and post-Red Giant Branch (post-RGB) binaries, provides valuable insights into the complex binary interaction process that concludes the red-giant phase of evolution in these systems. Additionally, it offers a novel opportunity to investigate the formation of second-generation planets within dusty discs surrounding evolved stars. We present a pilot multi-wavelength polarimetric imaging study of the post-AGB binary system IRAS\,08544-4431 using the European Southern Observatory-Very Large Telescope/SPHERE instrument. This study is focused on optical $V-$ and $I'-$band ZIMPOL data to complement near-infrared $H-$band IRDIS data presented previously. The study aims to investigate the dust scattering properties and surface morphology of the post-AGB circumbinary disc as a function of wavelength. We successfully resolved the extended disc structure of IRAS\,08544-4431, revealing a complex disc morphology, high polarimetric disc brightness (up to $\sim$1.5\%),  and significant forward scattering at optical wavelengths. Additionally, we found that the disc shows a grey polarimetric colour in both optical and near-infrared. The findings highlight similarities between post-AGB circumbinary discs and protoplanetary discs, suggesting submicron-size porous aggregates as the dominant surface dust composition, and indicating potential warping within the disc. However, further expansion of the multi-wavelength analysis to a larger sample of post-AGB binary systems, as well as high-resolution observations of dust continuum and gas emission, is necessary to fully explore the underlying structure of post-AGB circumbinary discs and associated physical mechanisms.

\end{abstract}




\section{Introduction}
\label{sec:intro}

The discovery of planets orbiting the white dwarf - M dwarf binary system NN Ser and subsequent studies have challenged traditional notions of planet formation \citep{Zorotovic2013A&A...549A..95Z, Marsh2016MNRAS.459.4518H}. It was shown that these planets did not form alongside their hosting binary system, but rather originated from the remaining matter bound to the binary after the common envelope stage \citep{Volschow2014A&A...562A..19V, Schleicher2015AN....336..458S}. Studying the formation and evolution of second-generation circumbinary discs around evolved binary stars, such as post-Asymptotic Giant Branch (post-AGB) and post-Red Giant Branch (post-RGB) binaries, offers opportunities to gain insights into the poorly understood binary interaction process that terminates the giant phase of evolution in these systems. It also offers a unique opportunity to investigate the potential formation of second-generation planets in dusty discs surrounding evolved stars.

Low-mass post-AGB stars undergo a rapid transition ($\sim 10^6$ years) between the AGB and the planetary nebulae stages \citep{VanWinckel2003ARA&A..41..391V, Kamath2015MNRAS.454.1468K, Bertolami2016A&A...588A..25M}. In the presence of a secondary companion \citep[likely to be a main-sequence star,][]{Oomen2018}, binary interaction processes lead to the formation of various structures, including i) a stable circumbinary disc \citep[e.g.,][]{Kluska2022}, ii) circumstellar disc around the secondary companion and iii) high-velocity outflows (referred to as jets) launched from the companion position \citep[e.g.,][]{Bollen2022arXiv220808752B, Verhamme2024A&A...684A..79V, DePrins2024arXiv240609280D}. Observational studies have revealed that post-AGB circumbinary discs exhibit Keplerian rotation and have a typical angular size of $\sim$0.5–1 arcsec ($\sim$100 to 500 AU), as deduced from CO position-velocity maps \citep[e.g.,][]{Gallardo_cava2021A&A...648A..93G}.

Radiative Transfer (RT) modelling efforts for circumbinary discs around post-AGB binaries \citep[e.g.,][]{Corporaal_IRAS08_2023A&A...671A..15C} show that the high-angular resolution interferometric data are well reproduced by passively irradiated disc models developed for protoplanetary discs (PPDs) around young stellar objects. Despite the significant difference in formation history, as well as the lifetime between post-AGB circumbinary discs \citep[with a lifetime of $\sim10^4-10^5$ years,][]{Bujarrabal2017A&A...597L...5B} and PPDs (with a lifetime of up to a few Myr), the two types of discs appear to be surprisingly similar in terms of IR excesses, dust disc mass \citep[$\sim10^{-3}M_\odot$,][]{Corporaal_IRAS08_2023A&A...671A..15C}, chemical depletion \citep{Kluska2022, Mohorian2024MNRAS.530..761M} and dust mineralogy \citep{Gielen2011A&A...533A..99G, Scicluna2020MNRAS.494.2925S}. 
Deep near-infrared (near-IR) interferometric studies spatially resolved the hot dust inner rim of the post-AGB discs, confirming that it is predominantly located at dust sublimation radius \citep[$\sim$6\,AU; e.g.,][]{Hillen2016, Kluska2019A&A...631A.108K}. These circumbinary discs are called 'full discs'. However, the recent study by \citet{Corporaal2023A&A...674A.151C} confirmed that similar to PPDs, we can define a subgroup of ‘transition' post-AGB discs with the dust inner rim up to 7.5 times larger than the theoretical dust sublimation radius. In PPDs, 'transition' disc characteristics are interpreted as a sign of cavities in the disc structure, which could reflect the presence of a giant planet carving a hole in the disc \citep{Maaskant2013A&A...555A..64M, Menu2015A&A...581A.107M, Booth2020MNRAS.493.5079B}.

Advanced direct imaging instruments, equipped with extreme adaptive optics (AO), provide access to scattered light from small dust grains in the disc surface layers \citep{Benisty2022arXiv220309991B}. By employing high-resolution polarimetric imaging with cutting-edge instruments like European Southern Observatory (ESO) Spectro-Polarimetric High-contrast Exoplanet REsearch instrument \cite[SPHERE,][]{Beuzit2019}, we can effectively suppress direct stellar light from the system. This allows us to unveil the extended disc morphology and extract crucial information about the dust properties encoded in both the intensity and degree of polarization of the scattered light \citep{Ma2023A&A...676A...6M}. Employing multi-wavelength polarimetric imaging, we can gain insights into the wavelength dependence of scattering parameters, including dust-scattering albedos, polarization intensity, scattering asymmetry, and disc colour. These parameters, compared with results of state-of-the-art dust models \citep[e.g.,][]{Tazaki2022A&A...663A..57T}, provide direct insights into the dust grain size, shape and composition in the surface layers of the circumstellar disc. 

AR Pup is the first post-AGB binary whose extended disc structure was resolved with SPHERE \citep{Ertel2019AJ....157..110E}. The extended disc showed several arc-like features extending perpendicular to the disc mid-plane and caused by a disc wind, outflows, or jets from the central binary. Since then, we have used SPHERE to resolve the extended disc of an additional 8 post-AGB binaries \citep{Andrych2023MNRAS.524.4168A}. This pilot survey revealed complex morphologies and also showed a large diversity in terms of disc size and orientation. The study also showed that 'full' disc systems have clearly resolved elliptical disc surfaces, in contrast to 'transition' discs, which exhibit more complex structures. Additionally, it indicated that post-AGB circumbinary discs exhibit a high azimuthal polarization of up to 2.2\% of stellar intensity, which is consistent with values found for PPDs. However, further investigation of the circumbinary disc structure and interpretation of the findings requires expanding the available data to include multi-wavelength polarimetric imaging observations.

In this paper, we present the first multi-wavelength polarimetric images of the disc around the post-AGB binary star IRAS\,08544-4431. In Section \ref{sec:target}, we present relevant observational details of our target collected from the literature. In Section \ref{sec:data}, we introduce our observing strategy and data reduction methodology. In Section~\ref{sec:analysis} we present the analysis of the VLT/SPHERE-ZIMPOL data and corresponding results. In Section~\ref{sec:discussion}, we discuss the effect of interstellar polarisation on the results and highlight similarities and differences between circumbinary discs around post-AGBs and PPDs. We present our conclusions in Section~\ref{sec:conclusion}.

\section{Target details}
\label{sec:target}

In this study we focus on the post-AGB binary star IRAS\,08544-4431, a luminous system \citep[$\sim 14000$\,L$_\odot $,][]{Oomen2019A&A...629A..49O} with well-established orbital parameters: an orbital period of $501.1\pm1.0$ days, a projected semi-major axis of $0.398\pm0.008$\,AU, an eccentricity of $0.20\pm0.02$, and a mass function of $0.033\pm0.002 M_\odot$ \citep{Oomen2018}. Spectroscopic observations also showed a mild depletion of refractory elements in the photosphere of the primary post-AGB star \citep[\texttt{[}Zn/Ti\texttt{]}=0.9,][]{Maas2005}. ALMA $^{12}$CO and $^{13}$CO observations revealed a relatively extended disc in the system, with the central part exhibiting purely Keplerian rotation and sub-Keplerian rotation towards the outer regions \citep{Bujarrabal2018}. 

Additionally, IRAS\,08544-4431 has been extensively studied in the near-IR and mid-IR with ESO/Very Large Telescope Interferometer (VLTI), showing a well-resolved inner disc rim aligned with the dust sublimation radius \citep{Hillen2016, Kluska2019A&A...631A.108K, Corporaal_IRAS08_2023A&A...671A..15C}. Therefore, the system was classified as a 'full' disc. However, it was also found that $\sim$15\% of the H-band flux comes from an over-resolved emission, only partially accounted for by current radiative transfer models, suggesting a missing component with temperatures ranging within 1400-3600 K \citep{Kluska2018A&A...616A.153K, Corporaal2021A&A...650L..13C, Corporaal_IRAS08_2023A&A...671A..15C}. 

Recent SPHERE/IRDIS $H$-band polarimetric imaging of IRAS\,08544-4431 have shown the low-inclination disc ($\sim20^\circ$) in the system, characterized by the polarized disc brightness of 0.58\% (in resolved emission) and 1.30\% (including the unresolved component) of the total intensity of the target \citep{Andrych2023MNRAS.524.4168A}. It also revealed complex disc surface structures, including two arc-like substructures alongside the main 'ring'. These results further support the non-axisymmetric nature of the system proposed by \citet{Corporaal2021A&A...650L..13C} and make IRAS\,08544-4431 a favourable target for probing the wavelength dependency of disc scattering and polarimetric parameters. 


\section{Data and Observations}
\label{sec:data}

In this study, we present multi-wavelength high-angular-resolution polarimetric observations of the circumbinary disc surrounding IRAS\,08544-4431. In this section, we provide an overview of our data and the methodology employed for data reduction. The data reduction process involves several steps, including polarimetric differential imaging (PDI), frame selection, correction of telescope effects and unresolved central polarization, deconvolution, and estimation of signal-to-noise ratio.

\subsection{Observations}
\label{observations}

The data were obtained with the extreme AO instrument \citep[SPHERE,][]{Beuzit2019} using the Zurich Imaging Polarimeter \citep[ZIMPOL,][]{Schmid2018A&A...619A...9S} in optical, as well as the Infra-Red Dual-beam Imaging and Spectroscopy camera \citep[IRDIS,][]{Dohlen2008} for the near-IR. The IRDIS observation results are adapted from \citet{Andrych2023MNRAS.524.4168A}. In Table~\ref{tab:disc_orient} we summarise the resulting disc parameters determined from IRDIS data as well as the results of this study (see Section~\ref{sec:analysis} for details).

We obtained the SPHERE/ZIMPOL data as part of ESO observational program 0101.D-0752(A) (PI: Kamath). Observations were carried out on April 22nd, 2018. To resolve the scattered light around the post-AGB binary star and ensure the best accuracy for the multi-wavelength characteristics, we used SPHERE/ZIMPOL in its polarimetric P1 mode without coronagraph. In this mode, the sky field is rotating with respect to the instrument and the telescope pupil so that instrumental effects average out to provide the highest polarimetric sensitivity and polarimetric calibration accuracy \citep{Schmid2018A&A...619A...9S}. In addition to the target data, we also obtained observations for the reference star HD 83878. This star was carefully selected to have similar brightness and colour to the science target, at a similar position as the corresponding science target when being observed after it. While the total intensity image of the scientific target can be used as a point-spread function (PSF) during the reduction process for some post-AGB systems \citep{Andrych2023MNRAS.524.4168A}, reference star observations give us more accurate results ensuring that the PSF is not broadened by scattered light from the circumbinary disc. 

We obtained the data in $V$-band ($\lambda_0 = 554\,\text{nm}$, $\Delta\lambda = 80.6\,\text{nm}$) and $I'$-band (labelled in ZIMPOL as 'Cnt820', $\lambda_0 = 817.3\,\text{nm}$, $\Delta\lambda = 19.8\,\text{nm}$)\footnote{The ZIMPOL camera includes three filters ('I\_PRIM', 'N\_I', and 'Cnt820') that correspond to the $I$-band and differ by the filter width. For the convenience of the reader, we will use the label $I'$-band to refer to 'Cnt820' throughout the paper.}. Due to the construction of the ZIMPOL camera, observations in both filters were taken simultaneously, enabling direct comparison of data during further analysis. The data were taken in polarimetric cycles, where one cycle includes four Stokes exposures for linear polarization measurements, with half-wave plate positions $\theta = 0^{\circ}$ and $45^{\circ}$ for Stokes Q and $\theta = 22.5^{\circ}$ and $67.5^{\circ}$ for Stokes U. In total 20 polarimetric cycles were taken with an exposure time of 80 seconds per cycle. All observational frames for IRAS\,08544-4431 contain a total number of counts of approximately $5.6 \times 10^{6}$ in $V$-band and $3.7 \times 10^{6}$ in $I'$-band, which correspond to non-saturated observations \citep{Schmid2018A&A...619A...9S}. Observations for the reference star HD\,83878 show a similar order of total counts—approximately $3.4 \times 10^{6}$ in $V$-band and $9.5 \times 10^{5}$ in $I'$-band.

\subsection{ZIMPOL Data reduction}
\label{sec:data_reduction}
We performed the data reduction (DR) using the data reduction pipeline SPHERE-DC \citep[SPHERE Data Centre,][]{Delorme2017sf2a.conf..347D} complemented with the sz-pipeline developed at ETH Zurich \citep{Schmid2018A&A...619A...9S} and our own $Python$ scripts. In the following sections, we briefly present the main steps involved in the DR procedure.

\subsubsection{Polarimetric differential imaging (PDI) reduction with SPHERE-DC}
\label{sec:pdi}

To perform the basic PDI reduction we used the High-Contrast Data Centre (HC-DC, formerly known as the SPHERE Data Centre) pipeline. This software is based on the SPHERE ESO pipeline, complemented by additional routines (e.g. improved centring routines, automatic frame sorting, analysis routines) to systematically process SPHERE data. The basic reduction steps include bias subtraction, flat fielding, frame centring, polarimetric combination and corrections for the polarimetric beam-shift effect. The pipeline then uses the double difference method to compute the classical Stokes vector components $Q$ and $U$ and the corresponding total intensities. While polarimetric normalization is commonly used in the data reduction of SPHERE/ZIMPOL data \citep[e.g.,][]{Hunziker2021A&A...648A.110H}, we avoided this calibration step due to detected unresolved polarization coming from the unresolved part of the circumbinary disc of IRAS\,08544-4431 (see Section~\ref{sec:unres_polarization} and Section~\ref{sec:interstellar} for more details). To achieve the highest accuracy for the barely resolved data we also performed additional reduction steps such as correction of variable telescope polarization with the parallactic angle and correction of PSF smearing effect (see Section~\ref{sec:telesc_polarization} and Section~\ref{sec:psf_smearing}).
We used custom Python scripts to compute the azimuthal Stokes parameters $Q_\phi$ and $U_\phi$ following the definitions of \citet{deBoer2020}, the total polarized intensity ($I_{\rm pol}$), and the Angle of Linear Polarization (AoLP). Positive $Q_\phi$ indicates linear polarization in the azimuthal direction, while a negative $Q_\phi$ corresponds to linear polarization in the radial direction. $U_\phi$ shows the linear polarization shifted by $\pm45^\circ$ from the azimuthal and radial directions.
AoLP reflects the orientation of the polarization vector in each region of the polarized image. Assuming that the polarimetric signal arises from the single scattering of stellar light on the disc surface and that the disc has low inclinations ($i\lesssim 40^\circ$), the polarization vectors are expected to align azimuthally with respect to the central star. Any other signals can then be attributed to noise or calibration errors \citep{Canovas2015A&A...582L...7C}.

Considering the limited angular size of the post-AGB circumbinary disc, adopting $Q_\phi$ as the measure of polarized flux offers significant advantages. This approach avoids bias effects from noisy observational data in the polarized intensity $I_{\rm pol}$, which is derived using the square root of the sum of $Q$ and $U$ squares \citep{Simmons1985A&A...142..100S}. To assess the validity of this approach for our data sample, we calculated the ratio of net positive $Q_\phi$ (azimuthally polarized signal) to $I_{\rm pol}$. We found that azimuthally polarized intensity accounts for more than 85\% of the total polarized intensity ($I_{\rm pol}$) in the $V$-band and over 90\% in the $I'$-band confirming the dominance of single scattering and justifying the use of $Q_\phi$ for the further analysis. 

\subsubsection{Frame selection}
\label{sec:frame_selection}

The PSF of SPHERE/ZIMPOL can change strongly within one night and under unstable weather conditions even from frame to frame, where each frame is the result of one full polarimetric cycle. This variability impacts polarimetric measurements, causing the measured polarization to also depend on the variable atmospheric conditions and the associated AO performance \citep{Tschudi2021A&A...655A..37T}. Therefore we investigated PSF variability for both science and reference observations for individual data frames.

\begin{figure} 
    \includegraphics[width=0.9\columnwidth]{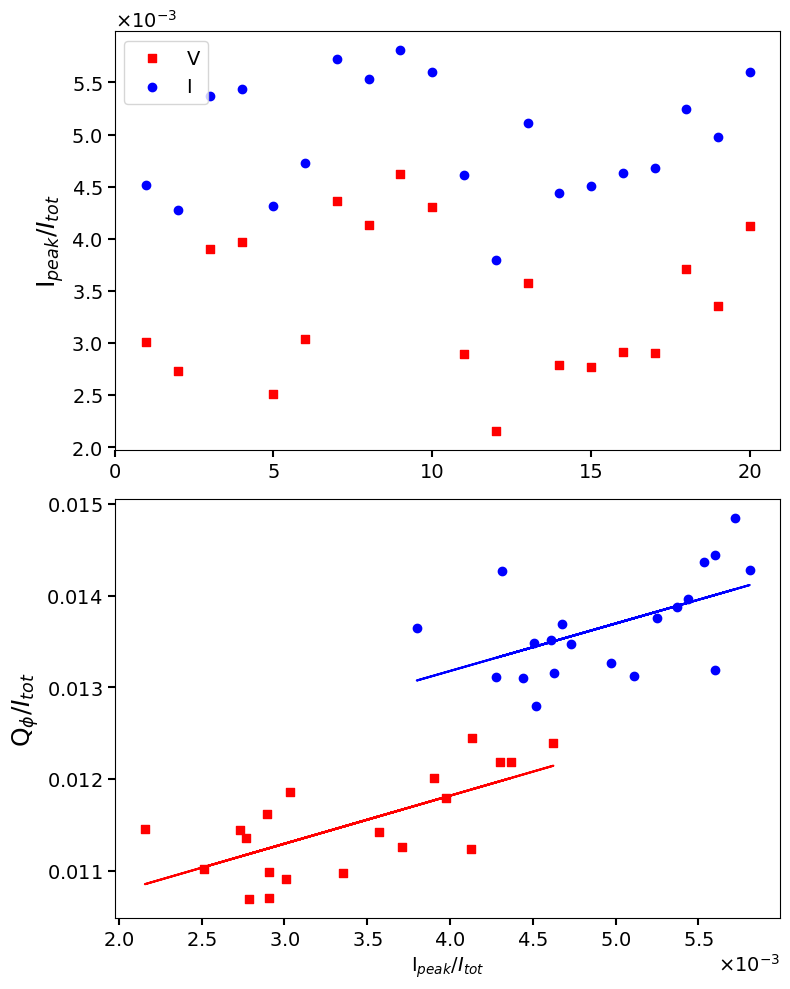}

    \caption{Frame-to-frame variation of normalized peak PSF intensity ($I_{\rm peak}/I_{\rm tot}$) as a simple measurement of observational Strehl ratio (top row), and correlation of normalized disc polarization ($Q_{\phi}/I_{\rm tot}$) with peak PSF intensity for IRAS\,08544-4431. The $V-$band (in blue) and $I'-$band (in red) data exhibit very similar behaviour, as they were captured simultaneously by the two arms of ZIMPOL, reflecting the same AO performance. See Section~\ref{sec:frame_selection} for more details.}
    \label{fig:peakI}
\end{figure}

To assess this PSF variability, we used the peak flux ($I_{\rm peak}$) normalised to the total intensity ($I_{\rm tot}$) of the target in a large aperture ($\sim3''$) as a simple measure of the Strehl ratio. Given that total intensity images $I_{\rm tot}$ of IRAS\,08544-4431 are primarily dominated by the post-AGB star (see Appendix~\ref{sec:scattered}), we used $I_{\rm tot}$ as the instrumental PSF taken simultaneously with the polarimetry data for this reduction step. The resulting frame-to-frame variation of the observational PSF is presented in the top row of Fig.~\ref{fig:peakI}. 

 Additionally, we explored the correlation between the normalized total peak flux ($I_{\rm peak}/I_{\rm tot}$) and the observed normalized disc polarization ($Q_{\phi}/I_{\rm tot}$). As anticipated, we observed a strong correlation in both the $V-$ and $I'-$bands (see bottom panel of Fig.~\ref{fig:peakI}).
 
 Finally, we used the normalized peak flux $I_{\rm peak}/I_{\rm tot}$ to select the top 75\% \footnote{The choice of the top 75\% of frames was based on a combination of factors, including the change of the peak intensity from frame to frame, visual inspection of the frames, and the total number of available frames. This value was selected to balance between excluding frames with poorer instrumental performance while still keeping enough frames to benefit from mean combining and significantly improve the signal-to-noise ratio.} of polarimetric frames for further reduction and analysis.

\subsubsection{Correction of telescope polarization}
\label{sec:telesc_polarization}

\begin{figure} 
    \includegraphics[width=1.0\columnwidth]{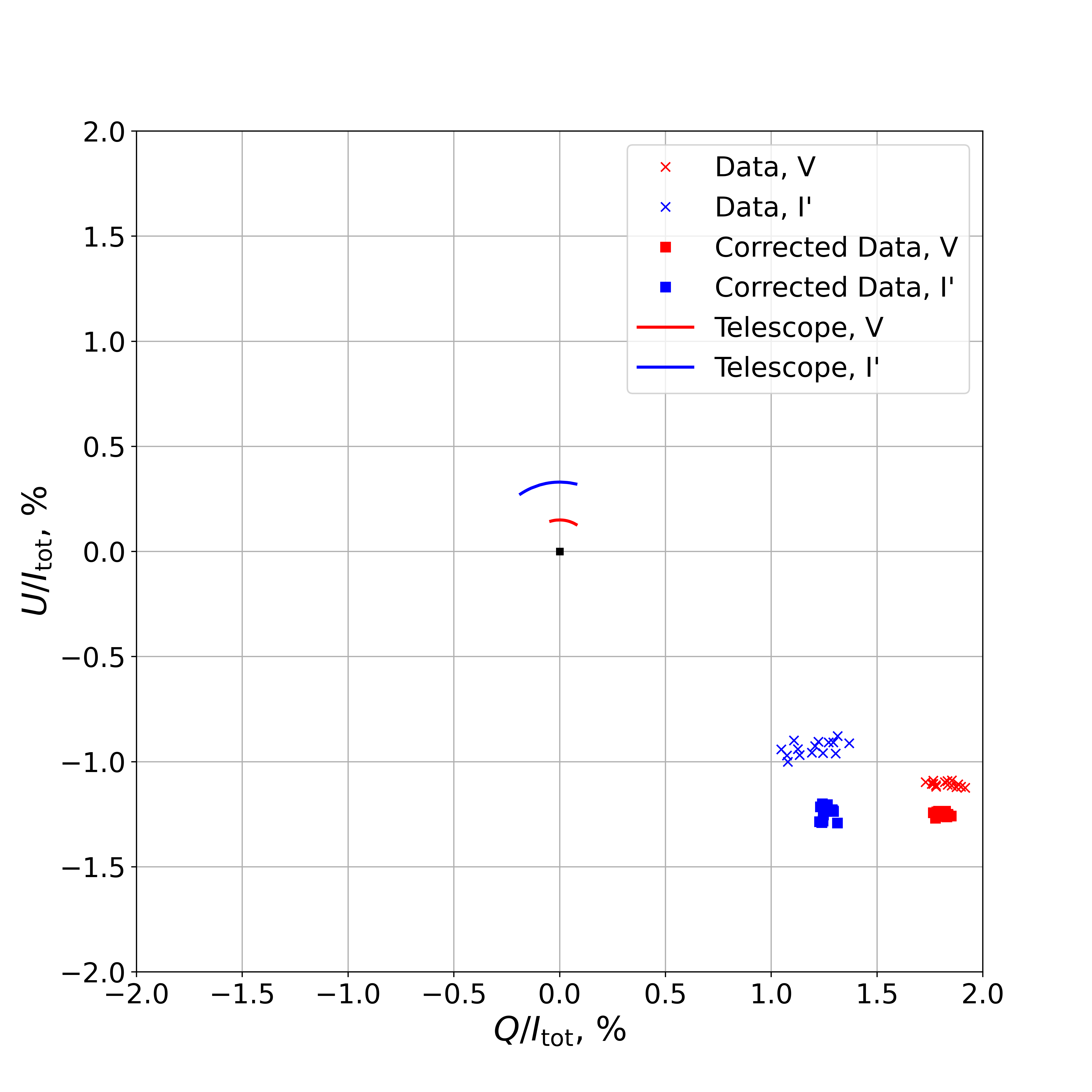}

    \caption{Measured fractional polarisation ($Q/I_{\rm tot}$ and $U/I_{\rm tot}$) of IRAS\,08544-4431 before (crosses) and after (squares) the correction of the telescope polarization (lines) in $V$ (red) and $I'$ (blue) bands. See Section~\ref{sec:telesc_polarization} for more details.}
    \label{fig:telesc_polarization}
\end{figure}

To correct our data for polarization induced by the instrument, we used the $sz$-pipeline developed at ETH Zurich. This pipeline, following the methodology outlined by \citet{Schmid2018A&A...619A...9S}, calculates the offset in telescope polarization and polarization position angle for the ZIMPOL instrument. As the telescope rotates relative to the sky during observations, the uncorrected fractional polarization ($Q/I_{\rm tot}, U/I_{\rm tot}$) typically forms a circular distribution in the $Q/I_{\rm tot}-U/I_{\rm tot}$ plane. Fig.\ref{fig:telesc_polarization} illustrates the measured fractional polarization before and after correction for each observation cycle, alongside the corresponding telescope polarization data for both $V-$ and $I'-$bands of IRAS\,08544-4431. The radius of the circle is determined by the telescope polarization, which was determined by \citet{Schmid2018A&A...619A...9S}. The position angle, on the other hand, is influenced by two factors: the parallactic angle and a specific offset unique to each band.

\subsubsection{Correction of the unresolved central polarized signal}
\label{sec:unres_polarization}

\begin{figure*} 
    \includegraphics[width=1.8\columnwidth]{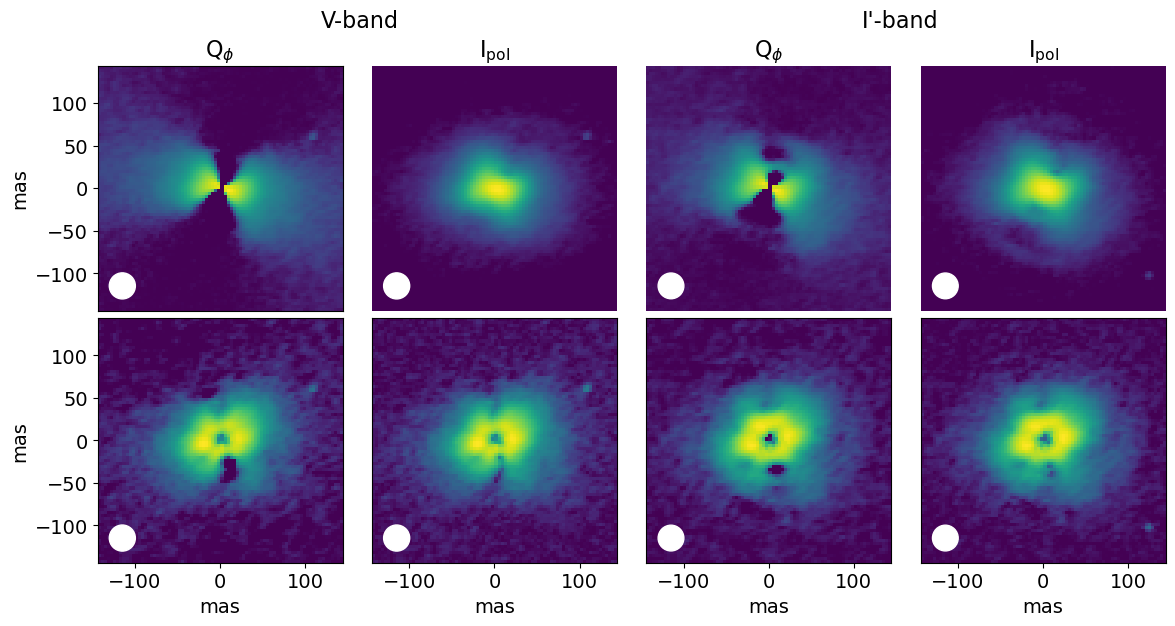}

    \caption{The polarized signal of IRAS\,08544-4431 before (top row) and after (bottom row) subtraction of the unresolved central polarization in $V$ (first and second column) and $I'$ (third and fourth column) bands. $Q_\phi$ represents azimuthal polarized intensity, while $I_{\rm pol}$ represents the total polarized signal. White circles in the left corner of each image represent the size of the resolution element (see Section~\ref{fig:snr}). All images are presented on an inverse hyperbolic scale and oriented North up and East to the left. See Section~\ref{sec:unres_polarization} for more details.}
    \label{fig:unres_polarization}
\end{figure*}

The reduced linearly polarized images of IRAS\,08544-4431 show features commonly attributed to the polarisation signal originating from unresolved components within the system \citep[e.g.,][]{Keppler2018A&A...617A..44K}. These features include a butterfly pattern in the central part of the $Q_\phi$, as well as a halo in the polarized total intensity image (see top row of Fig.\ref{fig:unres_polarization}). We note that similar unresolved polarized signal was also identified in $H-$band polarimetric observations of IRAS\,08544-4431 with SPHERE/IRDIS \citep{Andrych2023MNRAS.524.4168A}. Although the resolution of ZIMPOL camera in $V-$ and $I'-$bands is significantly better ($\sim 30$ mas, see Section~\ref{sec:snr}), it remains insufficient to resolve the inner rim of the circumbinary disc of IRAS\,08544-4431, located at approximately 5.5 mas from the central binary \citep{Kluska2018A&A...616A.153K}. 

To correct the data for the polarized signal from the bright but spatially unresolved central region, we implemented a method similar to that outlined in \citet{Holstein2020A&A...633A..64V}. Following the analysis performed in \citet{Andrych2023MNRAS.524.4168A}, we used a circular aperture with a radius of 3 pixels ($\sim11$ mas) centred on the stellar position to calculate the degree and angle of the unresolved central polarization (see Table \ref{tab:unresolved}). The 3-pixel aperture was chosen to provide a reliable correction for the unresolved polarized signal, while minimizing the contribution from the resolved signal. Subsequently, we subtracted this unresolved central polarized signal from the linearly polarized images $Q$ and $U$ and recalculated $Q_\phi$ and $I_{\rm pol}$. The resulting polarized images are presented in the bottom row of Fig.\ref{fig:unres_polarization}.

In addition, we note that employing this method corrects for the net linear polarization of the central source and therefore produces unrealistically low intensity for the central 5x5 pixel region of the $I_{\rm pol}$ and $Q_\phi$ images (see the bottom row of Fig.~\ref{fig:unres_polarization}). Even if there exists an intrinsic $Q_\phi$ component, it cannot be measured because of the lack of spatial resolution. Therefore, we exclude this region from further analysis and emphasize that this discrepancy is an artificial bias introduced during the reduction process.

\begin{table}
    \caption{Characteristics of the unresolved central polarisation.}
    \begin{center}
        \begin{tabular}{|c|c|c| }
            \hline
            band & DoLP, \% & AoLP, $^\circ$ \\
            \hline
            V& 0.95 $\pm$ 0.09& 160 $\pm$ 2\\
            I'&1.03$ \pm$ 0.06 &155 $\pm$ 1\\
         
            \hline
        \end{tabular}
    \end{center}
    \begin{tablenotes}
    
    \small
    \item \textbf{Notes:}  'DoLP' represents the degree of linear polarisation, 'AoLP' represents the predominant angle of linear polarisation for the unresolved central polarisation (see Section~\ref{sec:unres_polarization}). \\
    \end{tablenotes}
   
    \label{tab:unresolved}
\end{table}

\subsubsection{Correction of PSF smearing effect}
\label{sec:psf_smearing}

The correction of PSF smearing and polarimetric cancellation effects is known to be a crucial step in determining the polarized flux of a circumstellar disc \citep{Schmid2006A&A...452..657S, Tschudi2021A&A...655A..37T}. To properly account for the PSF convolution effects, one can define a parametric disc model and search for the initial polarized disc signal by comparing many PSF-convolved disc models with the observations. However, this approach requires a detailed model of the disc geometry that is not always known. 

Recently, \citet{Ma2024A&A...683A..18M} proposed a procedure that can be applied to a wide range of disc observations and allows to obtain corrected or intrinsic values for disc polarized intensity with relative errors of only about 10\% or even less. Following their methodology, we derived a two-dimensional correction map for the azimuthally polarized disc intensity $Q_\phi$ based on the observational PSF and disc orientation for IRAS\,08544-4431 \citep[inclination of 23$^\circ$ and position angle $\sim 121^\circ$ derived from $H-$band polarimetric imaging with SPHERE/IRDIS,][]{Andrych2023MNRAS.524.4168A}. We compared the resulting correction maps for the disc orientation (inclination and position angle) obtained from IR interferometry data \citep{Hillen2016, Corporaal_IRAS08_2023A&A...671A..15C} and did not find a significant difference in the results. To ensure accurate wavelength dependant comparison of SPHERE/ZMPOL and SPHERE/IRDIS data, we have also calculated similar correction map for the $H-$band SPHERE/IRDIS data that were already analysed and presented by \citet{Andrych2023MNRAS.524.4168A}. 

\begin{figure} 
    \includegraphics[width=1\columnwidth]{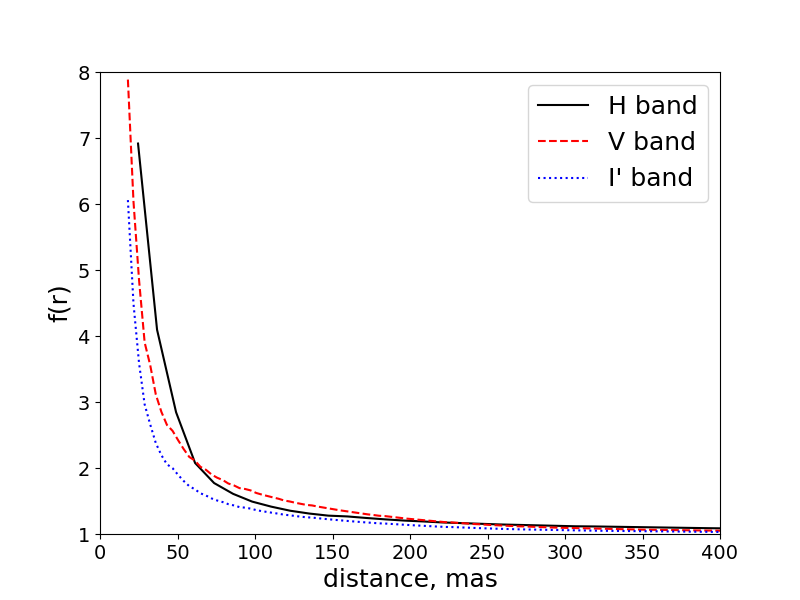}
    \caption{Example of calculated PSF correction coefficients in $H-$ (solid black line), $V-$ (dashed red line) and $I'-$bands (dotted blue line) using observational PSF and a wide elliptical ring model with inclination and position angle similar to IRAS\,08544-4431. See Section~\ref{sec:psf_smearing} for more details.}
    \label{fig:psf_smearing}
\end{figure}

In Fig.\ref{fig:psf_smearing} we present the radial curve $f(r)$ of the correction map for $V- , I-$ and $H-$bands. Similar to \citet{Ma2024A&A...683A..18M}, our result shows that the correction map has high values closer to the binary where the smearing by the PSF-peak already introduces strong polarimetric cancellation of a disc signal. This further highlights the importance of correction of PSF smearing and polarimetric cancellation effects for targets with small angular disc size. However, we note that this correction may introduce bias in the shape of resolved disc structures, primarily due to the extreme values of the correction map near the central binary. Therefore, while this correction was applied to restore the total polarimetric brightness of the disc, it was omitted during the determination of disc orientation and morphology.

\subsubsection{Estimation of the signal-to-noise ratio (SNR)}
\label{sec:snr}

\begin{figure} 
    \includegraphics[width=0.5\columnwidth]{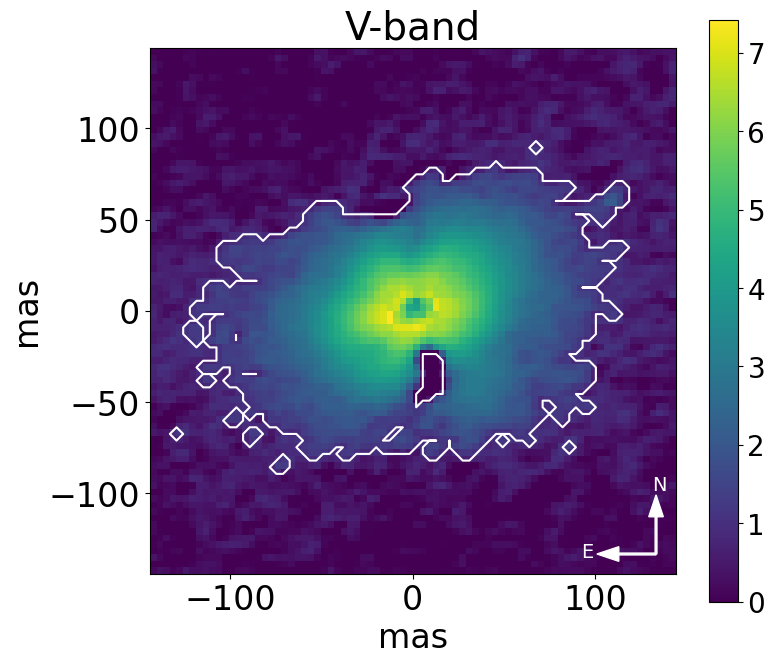}
    \includegraphics[width=0.5\columnwidth]{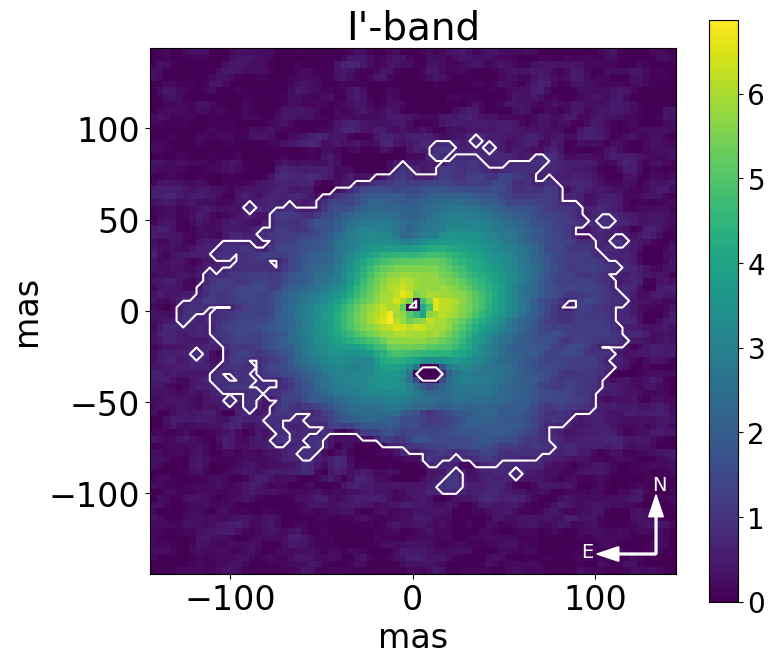}

    \caption{Statistically significant regions (SNR$\geq5$, marked with white contour) of the $Q_{\phi}$ images for IRAS\,08544-4431 in $V-$ and $I'-$bands. See Section~\ref{sec:snr} for more details.}
    \label{fig:snr}
\end{figure}

To determine the statistically significant region of the final reduced images, we used a similar approach to that applied for SPHERE/IRDIS observations of IRAS\,08544-4431 \citep{Andrych2023MNRAS.524.4168A}. In brief, we first determined the background noise for our datasets using the annulus of the image that did not contain any signal from the target and calculated the pixel-to-pixel root mean squared. Each pixel value of the azimuthally polarized image $Q_\phi$ was considered as the signal. We then used the calculated SNR to define the region of the $Q_\phi$ image with SNR $\geq5$, which reflects the statistically significant region of the polarimetric image for each observation (see Fig.\ref{fig:snr}). This region was used for further analysis. Additionally, we determined the resolution of our data to compare it with the expected performance of the SPHERE/ZIMPOL camera. Following the methodology outlined in \citet{Hunziker2021A&A...648A.110H}, we performed Gaussian fitting of the PSF intensity profile and used the full width at half maximum (FWHM) as the size of the resolution element. We found the resolution to be $\sim30$\,mas for both V- and I-bands, aligning closely with the anticipated values for the SPHERE/ZIMPOL instrument \citep{Schmid2018A&A...619A...9S}.

\subsubsection{Deconvolution}
\label{sec:deconvolution}

\begin{figure} 
     \includegraphics[width=0.5\columnwidth]{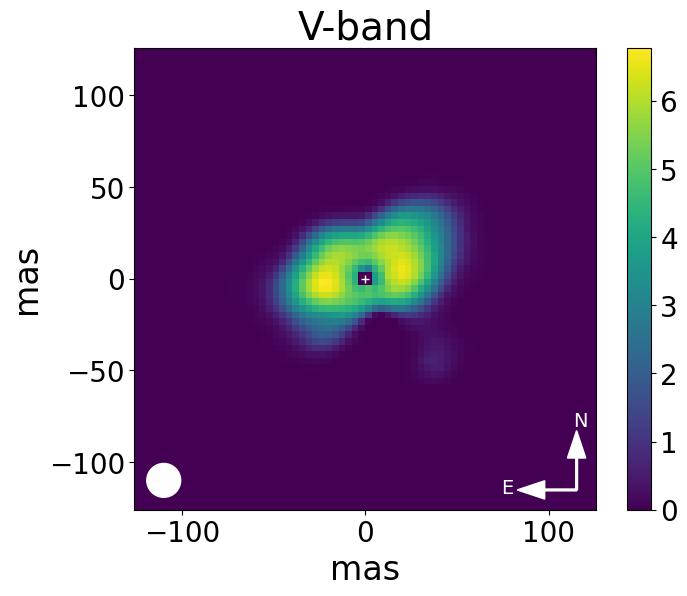} \includegraphics[width=0.5\columnwidth]{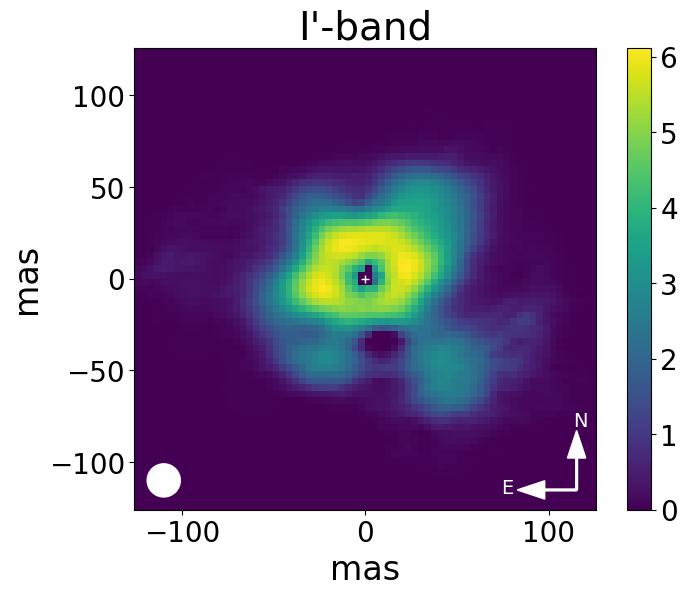}
    \caption{Deconvolved $Q_\phi$ images of IRAS\,08544-4431 in $V-$ and $I'-$bands. All images are presented on an inverse hyperbolic scale and oriented North up and East to the left. White circles in the left corner of each image represent the size of the resolution element after deconvolution. The low intensity of the central 5x5 pixel region of each image is a result of the correction for the polarization signal of the unresolved central source (Section~\ref{sec:unres_polarization}). See Section~\ref{sec:deconvolution} for more details.}
    \label{fig:deconvolution}
\end{figure}

To enhance the disc substructures, we deconvolved the resulting linearly polarized images with the reference PSF using the Richardson–Lucy deconvolution algorithm \citep{Richardson1972JOSA...62...55R, Lucy1974AJ.....79..745L}. We use the total unpolarized intensity image ($I_{\rm tot}$) derived from observations of a reference star as a PSF. We note that convergence of the deconvolution process was achieved within 25 iterations for both $V-$ and $I'-$bands, with pixel-to-pixel variation between neighbouring iterations reaching less than 1.5\%. We present the resulting deconvolved polarized images of IRAS\,08544-4431 in $V-$ and $I'-$bands in Fig.\ref{fig:deconvolution}. We also note that the final size of the resolution element after the deconvolution procedure is $\sim18$ mas.

\section{Analysis and Results}
\label{sec:analysis}

In this section, we present the analysis of reduced linearly polarized images and discuss the results for IRAS\,08544-4431. The main analysis procedure includes investigating the fractional polarisation, determining the disc orientation, defining the relative disc brightness in polarized light, and characterising the wavelength-dependent polarized intensity and structure of the circumbinary disc of IRAS\,08544-4431. As we noted in Section~\ref{sec:snr}, we focus on the region of the azimuthally polarized image with a signal-to-noise ratio $\geq5$ (see Fig.\ref{fig:snr}).

\subsection{Investigating  variations in aperture polarimetry}
\label{sec:aper_pol}

To explore potential geometric variations in the circumbinary disc, we examined how fractional polarization ($Q/I_{\rm tot}$, $U/I_{\rm tot}$) changes with increasing distance from the central binary. This approach also provides insights into the geometry of the unresolved disc component, which is the source of the unresolved central polarization detected in the system (see Section \ref{sec:unres_polarization}). If the disc has significant tearing or misalignment \citep[e.g.,][]{Perez2018ApJ...869L..50P} it would appear as rapid shifts in signal location within the $Q/I_{\rm tot}-U/I_{\rm tot}$ plane.

\begin{figure} 
    \includegraphics[width=0.9\columnwidth]{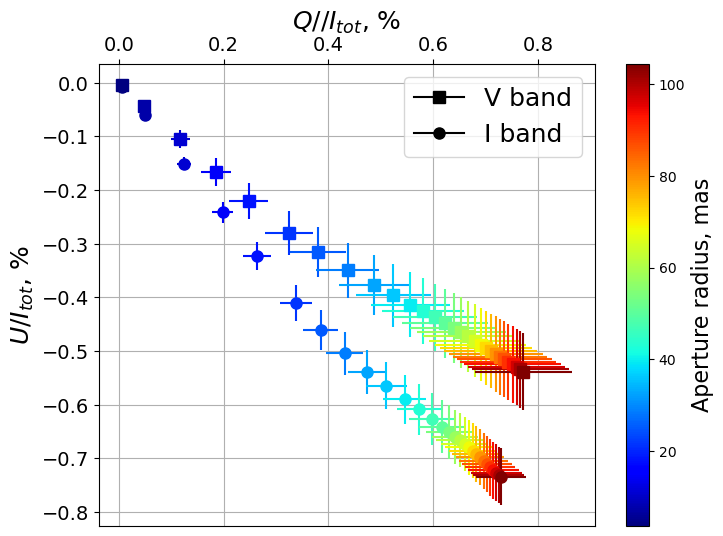}

    \caption{Fractional polarization $Q/I_{\rm tot}$ and $U/I_{\rm tot}$ for a gradually enlarging aperture for both $V-$ and $I'-$bands. See Section~\ref{sec:aper_pol} for more details.}
    \label{fig:aper_pol}
\end{figure}

We measured total fractional polarization $Q/I_{\rm tot}$ and $U/I_{\rm tot}$ in circular apertures with a gradually increasing radius ranging from 0.004''  to 0.11'' with step of 0.0036'' (1 pixel). The resulting values (see Fig.\ref{fig:aper_pol}) show a smooth and continuous change with aperture size, which is expected for the circumbinary disc without significant tearing and misalignment. While we note that interstellar polarization contributes to the unresolved central polarization, we expect that its impact on the measurement for IRAS\,08544-4431 is not significant, which is further discussed in Section \ref{sec:interstellar}.

\subsection{Determination of disc orientation}
\label{sec:disc_orient}

The final reduced $Q_\phi$ images reveal a clear bright 'ring' evident in both $V-$ and $I'-$band observations (see Fig.~\ref{fig:snr}), similar to our findings from SPHERE/IRDIS $H-$band polarimetric imaging of IRAS\,08544-4431 \citep{Andrych2023MNRAS.524.4168A}. While we refer to the resolved structure as a 'ring', it is important to clarify that the 'inner rim' in these images does not correspond to the physical dust inner rim in the IRAS\,08544-4431 circumbinary disc, which is too small to be resolved by the SPHERE instrument. An unrealistically low intensity at the centre of the $Q_\phi$ image is caused by the correction of the unresolved polarization of the central source during the DR (see Section \ref{sec:unres_polarization}). Therefore, the bright 'ring' represents the closest resolved section of the circumbinary disc to the central binary and serves as a valuable indicator for estimating the position angle (PA) and inclination of the disc.

Following the methodology discussed in \citet{Andrych2023MNRAS.524.4168A}, we used the deconvolved  $Q_\phi$ images to estimate the orientation of the resolved disc surface. This involved fitting an ellipse to the positions of peak brightness along the 'ring', from which we calculated the lengths of the major and minor axes, as well as the corresponding PA\footnote{The PA was defined as rising counterclockwise from the vertical axis (North) to the first principal radius (major axis). We note that the ellipse fitting procedure allows the $180^{\circ}$ ambiguity in the disc PA on the sky.} of the disc.

Radiative transfer modelling of the near-IR interferometric dataset for IRAS\,08544-4431 indicates that the dust inner rim aligns with the dust sublimation radius, rather than the dynamical disk truncation caused by the inner binary \citep{Kluska2018A&A...616A.153K}. Therefore, despite the known eccentricity of the post-AGB binary orbit \citep[$0.20\pm0.02$,][]{Oomen2018}, we can assume the disk to be circular and interpret the visible elliptical disc shape (see Fig.~\ref{fig:deconvolution}) as resulting from the projection of the circular disc onto the field of view. Hence, we can estimate the corresponding inclination that would cause such an effect using:
\begin{equation}
    \cos{i}=(\frac{b}{a}),
	\label{i}
\end{equation}
where $a, b$ are major and minor half axes of the ellipse.

Additionally, we determined the eccentricity $e$ of the 'ring' using:
\begin{equation}
    e=\sqrt{1-\frac{b^2}{a^2}}.
	\label{e}
\end{equation}

\begin{figure} 
    \includegraphics[width=0.5\columnwidth]{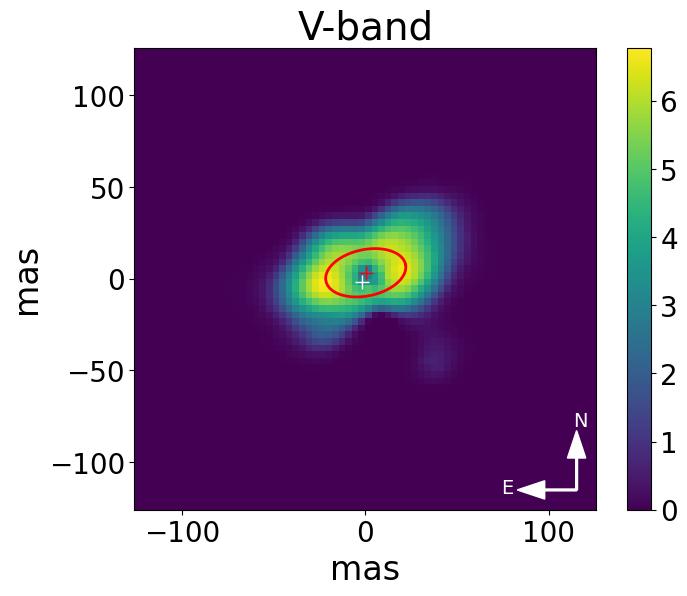}
    \includegraphics[width=0.5\columnwidth]{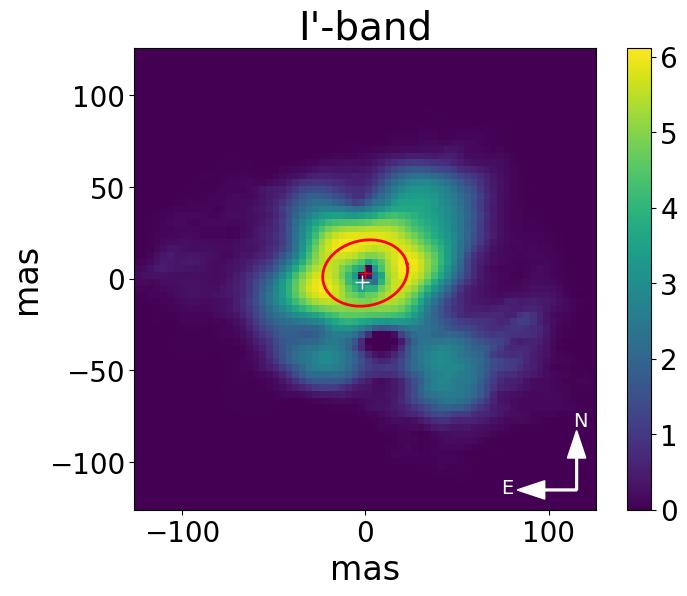}

    \caption{Disc orientation results based on the Q$_\phi$ polarized intensity images of IRAS\,08544-4431 (see Section~\ref{sec:disc_orient} for more details). The red ellipses illustrate the most plausible PA and inclination of discs. The red cross in the centre of the images represents the centre of the fitted ellipse, while the white cross represents the position of the binary. The low intensity of the central 5x5 pixel region of each image is a reduction bias caused by over-correction of the unresolved central polarization (Section~\ref{sec:unres_polarization}). Note: all images are presented on an inverse hyperbolic scale.}
    \label{fig:disc_orient}
\end{figure}

The most plausible position and orientation of circumbinary discs surface resolved in $V$ and $I'-$bands are presented in Fig.~\ref{fig:disc_orient} and Table~\ref{tab:disc_orient}. While correction of unresolved central polarization (see Section \ref{sec:unres_polarization}) may slightly circularize the resolved disc surface by over-subtracting fainter pixels, it is important to note that the AoLP of unresolved central polarization does not align with the minor axis of the resolved disc surface. Thus, this correction does not impact the calculated inclination and eccentricity values for IRAS\,08544-4431. We also note that the position angle of the disc matches for both $V-$ and $I'-$band observations presented in this paper, as well as in the $H-$band determined previously in \citet{Andrych2023MNRAS.524.4168A}. However, we observe that the inclination and eccentricity values increase with decreasing wavelength. We discuss the potential causes of such wavelength-dependent effects in Section~\ref{sec:discus_morph}.

    \begin{table}
        \caption{Summary of derived properties of IRAS 08544-4431 circumbinary disc in V, I and H-bands.}
        \begin{center}
            \begin{tabular}{ |l|c|c|c|c|c|c|}
            \hline
            band& \begin{tabular}[c]{@{}c@{}}
            $a$ \\ {[}mas{]}\end{tabular} &  \begin{tabular}[c]{@{}c@{}}$b$\\ {[}mas{]}\end{tabular}&
            \begin{tabular}[c]{@{}c@{}}
            $i$ \\  {[}$^\circ${]}
            \end{tabular} & \begin{tabular}[c]{@{}c@{}}
            $PA$ \\  {[}$^\circ${]}
            \end{tabular}& $e$ &\begin{tabular}[c]{@{}c@{}}
            $Q_{\phi}/I$ \\  {[}$\%${]}
            \end{tabular} \\
         
            \hline
            V & 22.2$^{+4}_{-3}$& 13$^{+2}_{-2}$& 55$^{+10}_{-12}$ & 102$^{+10}_{-10}$ & 0.82& 1.33$\pm$0.23\\
            I' & 23$^{+1}_{-1}$& 18$^{+1}_{-1}$& 40$^{+5}_{-6}$ & 102$^{+8}_{-8}$ & 0.65&1.46$\pm$0.14\\ 
            H$^*$ &38.6$^{+1.2}_{-1.2}$ & 35.5$^{+1.5}_{-1.4}$& 23$^{+7}_{-15}$ & 121$^{+27}_{-30}$ & 0.39& 1.03$\pm$0.2$^{\dag}$\\
            \hline
        \end{tabular}
    \end{center}
    \begin{tablenotes}
     \small
    \item \textbf{Notes:} $a$ and $b$ represent the major and minor half-axes of the disc in mas, $i$ indicates the inclination, $e$ represents the eccentricity. The position angle ($PA$) is presented in degrees and rises counterclockwise from the vertical axis (North) to the first principal radius (major axis).\\
    $^*$ indicates that results for the SPHERE/IRDIS $H-$band observations of IRAS\,08544-4431 were adopted from \citet{Andrych2023MNRAS.524.4168A}.\\
    $^{\dag}$ indicates that while the disc polarised brightness discussed in \citet{Andrych2023MNRAS.524.4168A} wasn't adjusted for the PSF smearing effect (see Section~\ref{sec:psf_smearing}), the value presented here has been corrected for this effect. This correction ensures a consistent comparison with disc polarised brightness in both $V-$ and $I-$bands presented in this study.\\
    
    \end{tablenotes}
       
        \label{tab:disc_orient}
    \end{table}

\subsection{Exploring the extended disc morphology}
\label{sec:morphology}

In this section, we investigate the extended disc morphology of IRAS\,08544-4431 spanning from $\sim20$ to 100 mas. To do this, we measure the polarized disc brightness relative to the total intensity of the system and analyse disc brightness profiles in both $V$ and $I'$-bands.

\subsubsection{Measuring the polarized disc brightness}
\label{sec: polarized_bright}

To estimate the polarized disc brightness relative to the total intensity of the system, we calculated the ratio of the resolved polarized emission from the disc to the total unpolarized intensity of the target. The resolved polarized emission was measured within the area, constrained by the SNR $\geq5$ (see Fig. \ref{fig:snr} and Section \ref{sec:snr} for details). Additionally, the total stellar intensity was measured within a 3" area centered on the central post-AGB source. To minimise bias, we used the total polarized and unpolarized images before deconvolution with the PSF, while accounting for the PSF smearing effect (see Section \ref{sec:deconvolution} and \ref{sec:psf_smearing}). We note that the resulting values for polarized disc brightness represent a lower limit due to the partial subtraction of the disc signal during data reduction (see Section~\ref{sec:unres_polarization}). The resulting polarized disc brightness ratios for both $V-$ and $I'-$bands are presented in Table~\ref{tab:disc_orient}. We also note that the estimated polarized disc brightness is independent of the stellar luminosity and distance to the post-AGB binary, as both the stellar light and the disc intensity exhibit similar scaling with distance.

In addition, we calculated the polarized contrast of our disc using the methodology proposed by \citet{Garufi2014A&A...568A..40G, Garufi2017A&A...603A..21G}. This method involves calculating the ratio of the observed polarized flux at a specific location to the net stellar flux incident on that region of the disc (i.e., the total stellar flux diluted by the distance). The measurement is conducted along the disc's major axis to minimise dependence on disc inclination and the unknown scattering phase function, enabling direct comparison with other systems. To determine the polarized contrast of IRAS\,08544-4431, we utilized the $Q_\phi$ signal within the resolved bright 'ring' to avoid potential bias from the asymmetric extended signal. The resulting value was averaged for opposite sides of the 'ring'. We calculated the polarized contrast of IRAS\,08544-4431's circumbinary disc to be $22\pm6 \times 10^{-3}$ in the V-band and $23\pm8 \times 10^{-3}$ in the I-band.

\subsubsection{Calculating disc brightness profiles}
\label{sec:profiles}

To investigate the complexity of the resolved disc surface, we calculated brightness profiles. These profiles reveal the spatial distribution of the polarized intensity of the disc. We calculated three types of brightness profiles: linear, azimuthal and radial following the procedure presented in \citet{Andrych2023MNRAS.524.4168A}. 

\begin{figure*} 
     \includegraphics[width=1\columnwidth]{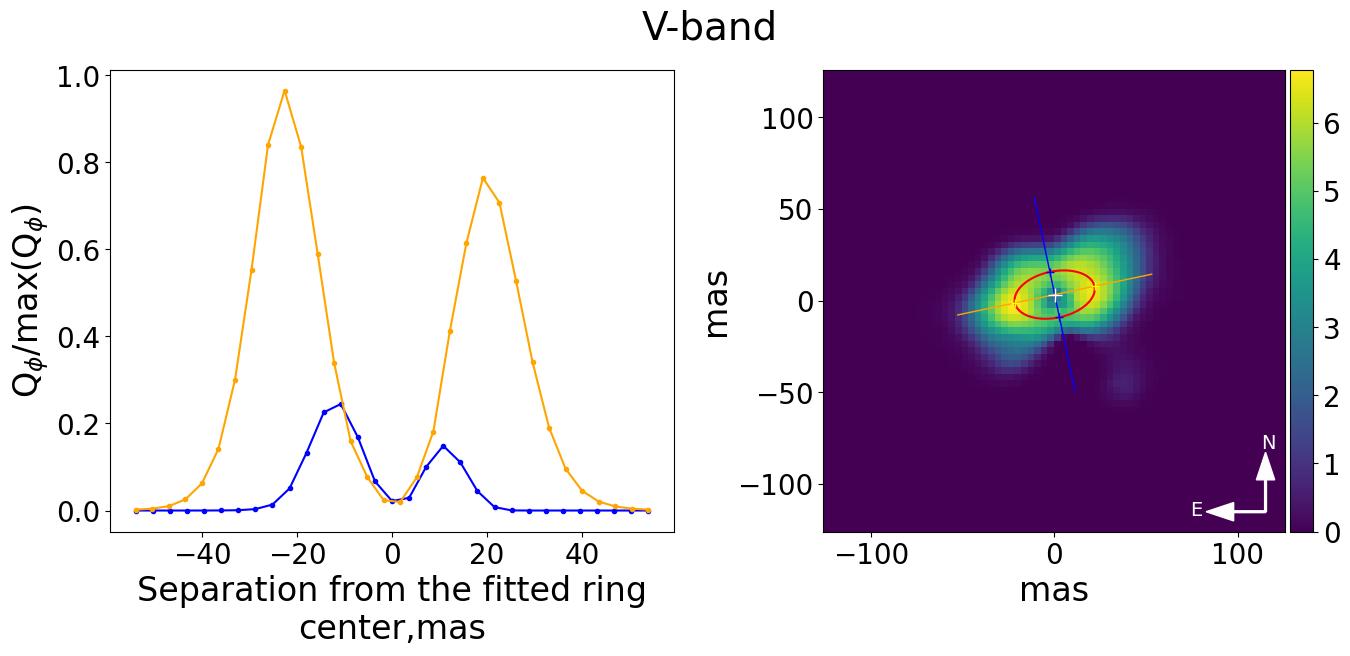}
     \includegraphics[width=1\columnwidth]{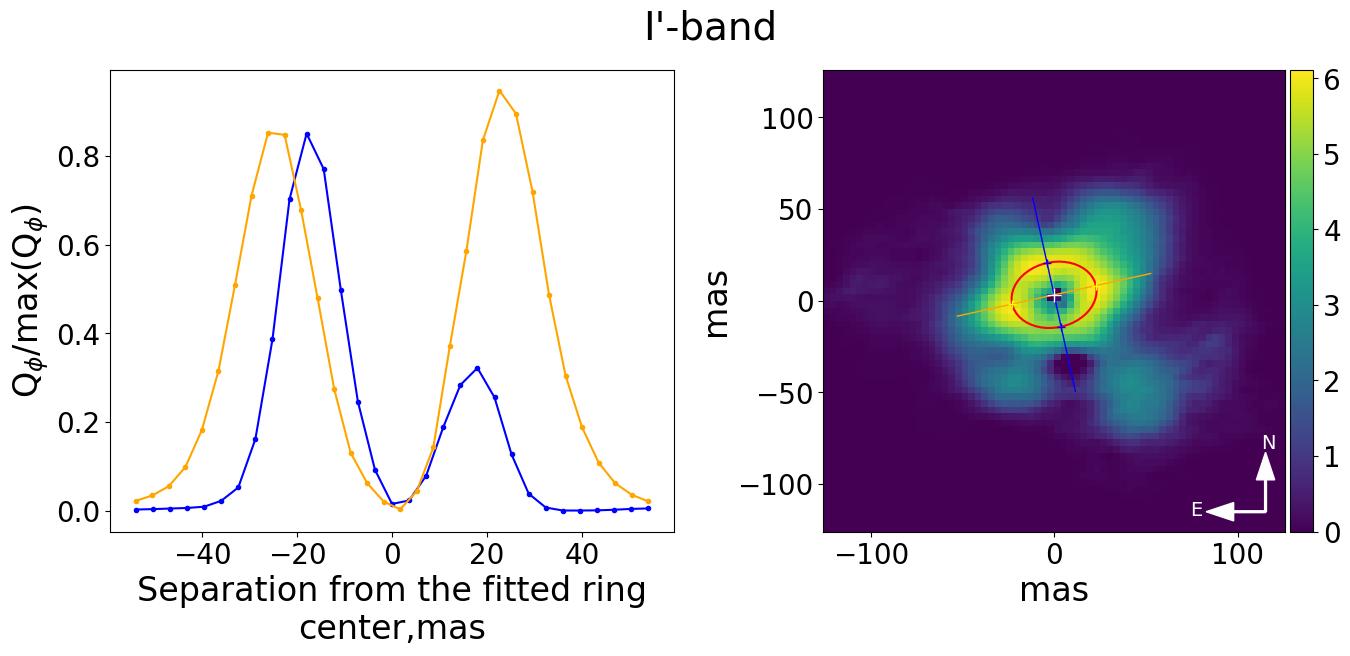}
    
    \includegraphics[width=1\columnwidth]{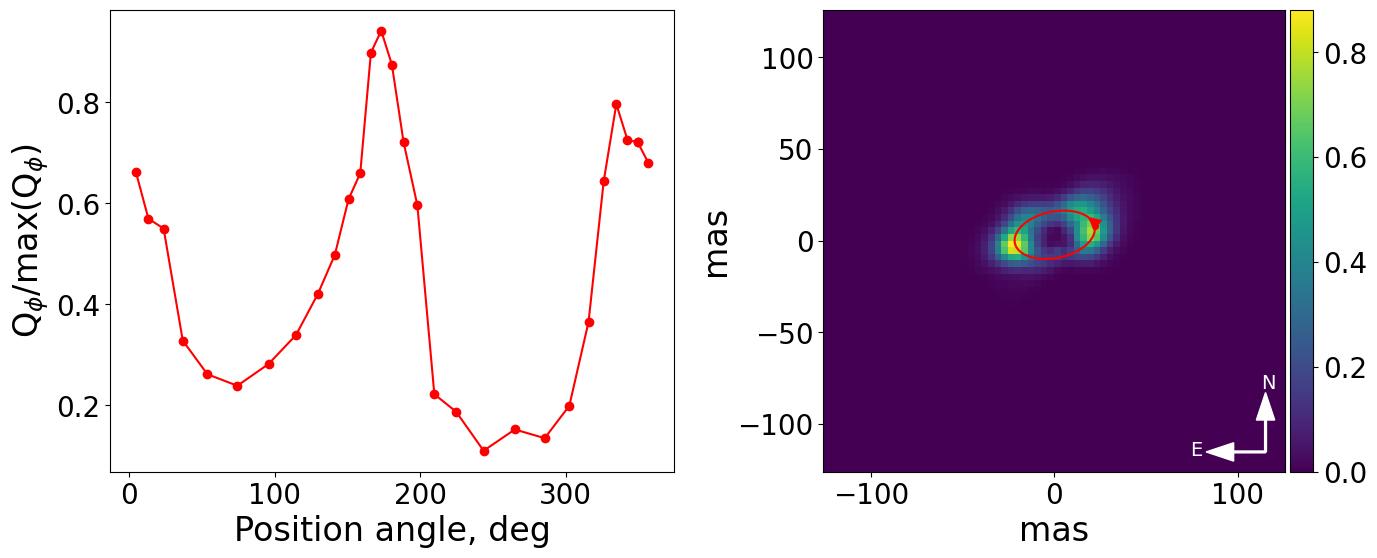}
     \includegraphics[width=1\columnwidth]{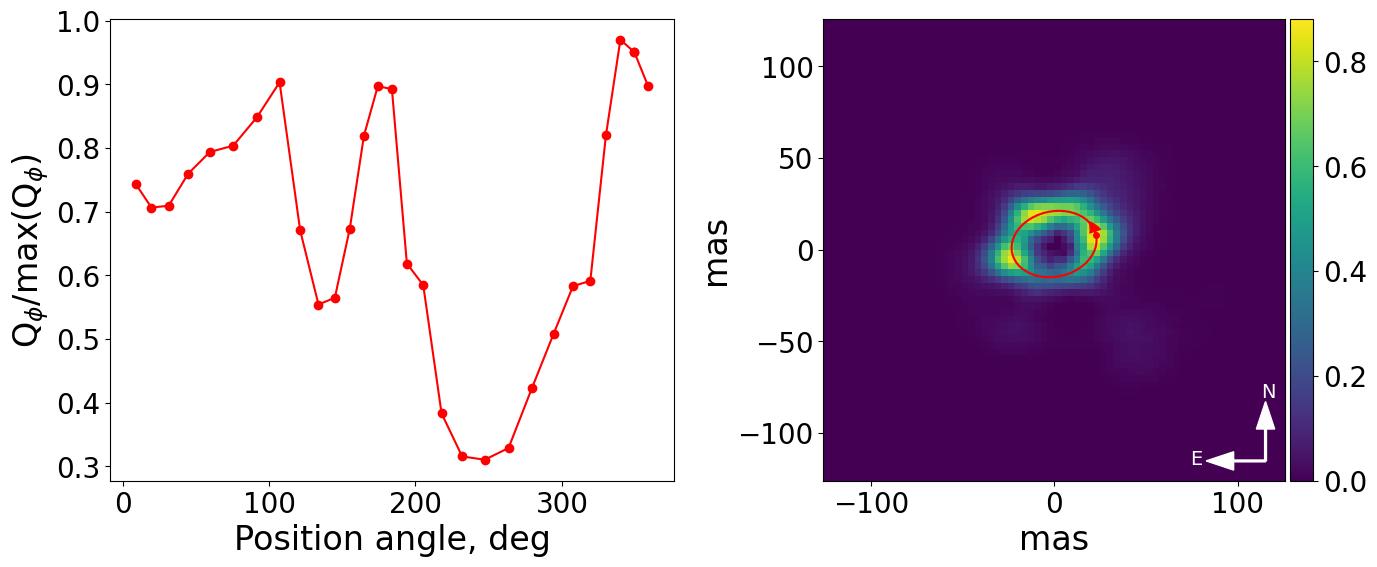}
     \includegraphics[width=1\columnwidth]{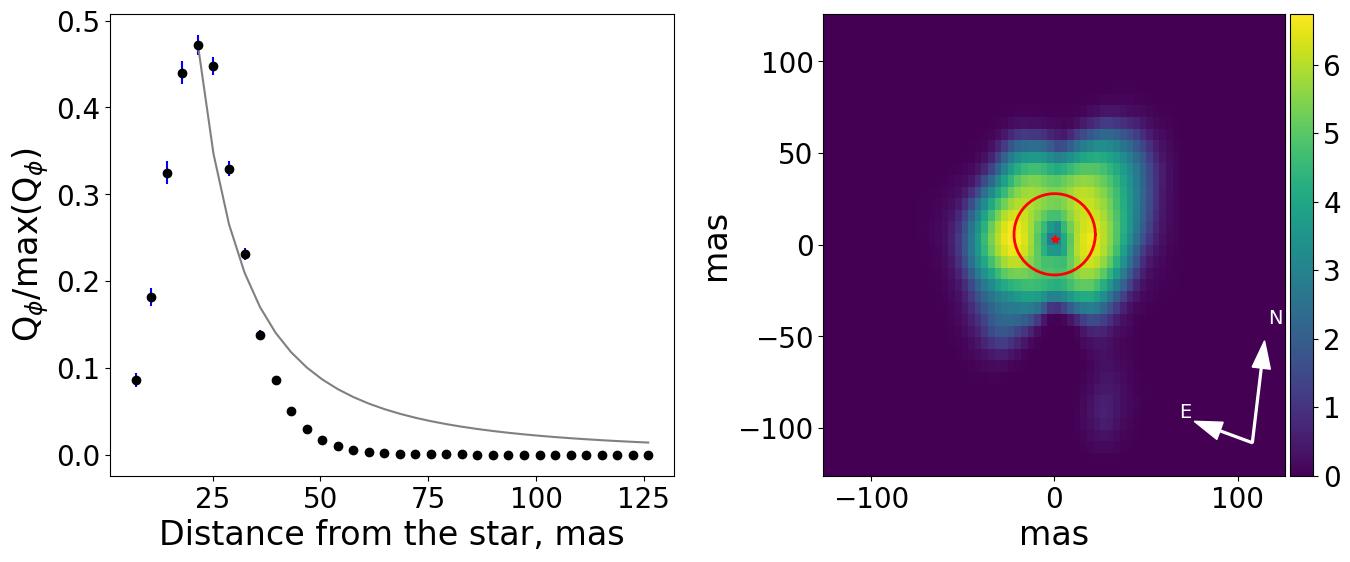} 
    \includegraphics[width=1\columnwidth]{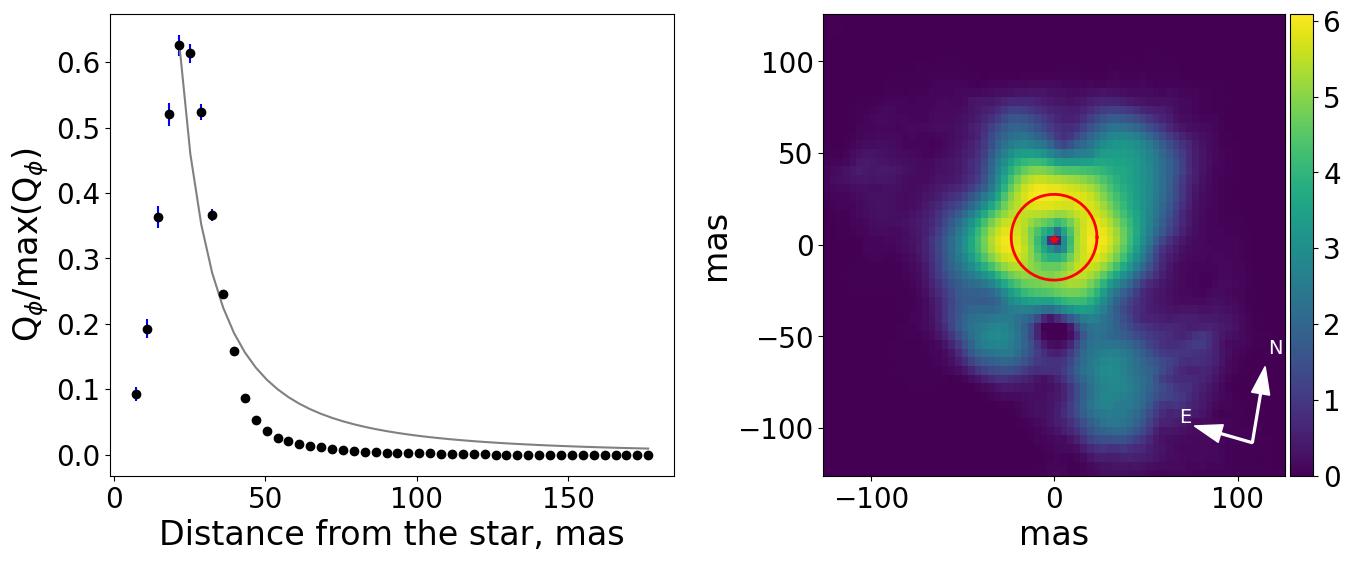}

    \caption{Brightness profiles for IRAS\,08544-4431 in $V$ (left panel) and $I'$ (right panel) bands (see Section~\ref{sec:profiles}) with corresponding  $Q_\phi$ images. In each panel, the left image displays the brightness profile, while the right image presents the corresponding $Q_\phi$ image. The top row shows linear brightness profiles of the $Q_\phi$ image along the major and minor axes of the fitted ellipse. The middle row represents the azimuthal brightness profiles of the $Q_\phi$ image. The red dot and arrow in the corresponding $Q_\phi$ image mark the starting point and direction of the azimuthal brightness profile calculation. The bottom row showcases radial brightness profiles of the deprojected $Q_\phi$ image. In the radial brightness profile plots, grey solid lines are added to indicate a r$^{-2}$ drop-off, expected from a scattered light signal due to the dissipation of stellar illumination. $Q_\phi$ images are presented on an inverse hyperbolic scale, with the middle-row images additionally normalized to the maximum $Q_\phi$ for clearer representation. The low intensity of the central 5x5 pixel region of each  $Q_\phi$ image is a reduction bias caused by correction of the unresolved central polarization (see Section~\ref{sec:unres_polarization}).}
    \label{fig:profiles}
\end{figure*}

The linear brightness profiles reflect the symmetry of the disc along its major and minor axes (see Section~\ref{sec:disc_orient} for the details on determining disc orientation). These profiles are shown in the top row of Fig.~\ref{fig:profiles}. We note that both $V-$ and $I'-$band linear brightness profiles along the minor axis of the resolved disc surface reveal substantial asymmetry, with the northern side of the disc appearing brighter than the southern side. Additionally, the $V-$band polarimetric image also exhibits slight but statistically significant asymmetry between the eastern and western sides of the disc. 

The azimuthal brightness profile reflects the asymmetries along the closest resolved part of the circumbinary disc to the central binary ('ring'). This profile was computed using Bi-linear interpolation of the intensity of the four nearest pixels at evenly spaced intervals along the ellipse fit in the $Q_\phi$ image (see Section~\ref{sec:disc_orient} for details). Starting from the eastern end of the major axis and proceeding counterclockwise, the profile traces the variations in azimuthal brightness. The resulting brightness profiles are presented in the middle row of Fig.~\ref{fig:profiles}. We note that both $V$ and $I'-$band azimuthal brightness profiles exhibit brightness peaks along the major axis of the resolved disc surface ($\sim0^\circ$ and $180^\circ$ in positional angle along the disc surface). This phenomenon is commonly attributed to polarization being most effective at scattering angles close to 90$^\circ$, which aligns with the sides of the major axis for inclined discs. However in $I'-$band azimuthal brightness profile, we also observe an increase in disc brightness from $20^\circ$  to $90^\circ$  in positional angle along the disc surface. We argue that this increase is caused by the forward scattering peak, suggesting that the northern part of the disc is inclined closer to the observer. Additionally, a similar phenomenon is observed in the $V-$band azimuthal brightness profile, although the asymmetry is less pronounced.

The radial brightness profile describes the variation of the scattered polarized emission with the distance from the central binary. To construct this profile, we calculated the mean brightness per pixel for radially tabulated annuli of the $Q_\phi$ images in both $V-$ and $I'-$bands. Considering the impact of disc orientation on the radial intensity distribution, we first reconstructed how the disc might look from the face-on (projected) position based on the estimated inclination of the resolved disc surface for both $V-$ and $I'-$bands (see Section~\ref{sec:disc_orient} for details on the inclination and \citet{Andrych2023MNRAS.524.4168A} for more information on the methodology). The deprojected $Q_\phi$ images were then subdivided into radially tabulated annuli, with the width of the annuli increasing proportionally to $\sqrt{r}$, where $r$ is the corresponding radius (similar to the methodology used by \citet{Avenhaus2018ApJ...863...44A} for PPD polarimetric observations). The resulting radial brightness profile was plotted as the mean brightness per pixel of an annulus against the separation from the post-AGB binary. These calculations were performed only for the statistically significant region of each $Q_\phi$ image based on the SNR (see Section~\ref{sec:snr}).

To estimate the noise, we used the variance in the same annuli of the $U_\phi$ image as $U_\phi$ is mainly devoid of signal, with the net $U_\phi$ not exceeding 5\% of the $Q_\phi$. Although this method may overestimate the noise if any astrophysical signal is present in the $U_\phi$ image, it still allows us to establish an upper limit. The resulting deprojected $Q_\phi$ images and radial brightness profile of IRAS\,08544-4431 are presented in the bottom row of Fig.\ref{fig:profiles}. We note that in both $V-$ and $I'-$bands the radial brightness profiles show only the main peak that corresponds to the bright 'ring'.

\subsubsection{Detection of  substructures}
\label{sec:substr}

The SPHERE/IRDIS near-IR study revealed asymmetrical substructures in the IRAS\,08544-4431 circumbinary discs \citep{Andrych2023MNRAS.524.4168A}. To investigate these substructures, we followed a similar qualitative analysis methodology to detect and characterize potential substructures in the circumbinary disc in $V-$ and $I'-$band polarimetric imaging. To define any real features in our Q$_{\phi}$ images (such as arcs or gaps), we combined the estimated SNR, the orientation of the polarization vector (AoLP), and the brightness of the disc substructures.

\begin{figure} 
    \includegraphics[width=0.5\columnwidth]{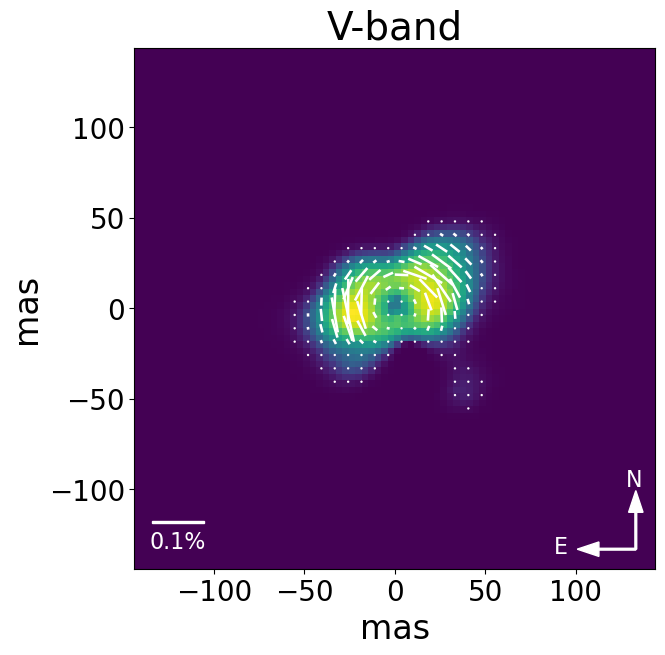}
    \includegraphics[width=0.5\columnwidth]{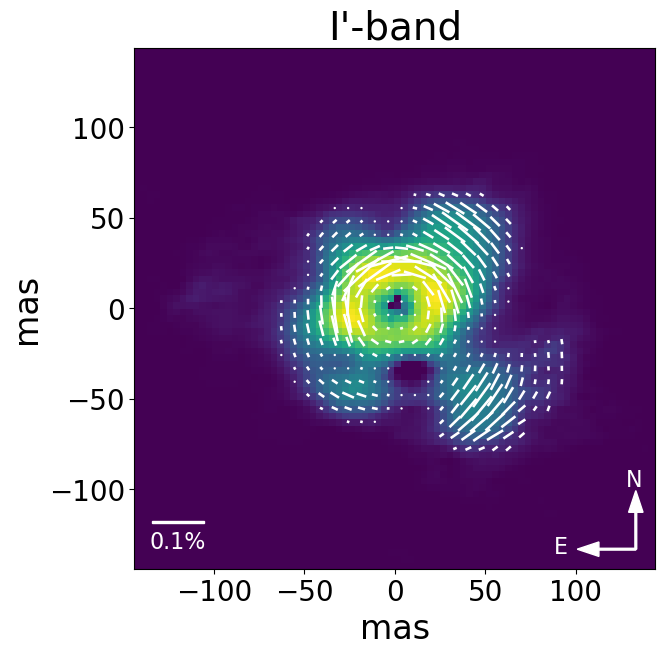}
    \includegraphics[width=0.5\columnwidth]{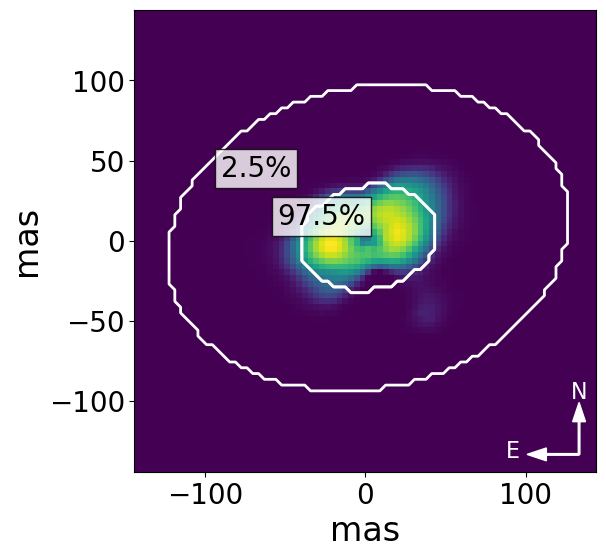}
    \includegraphics[width=0.5\columnwidth]{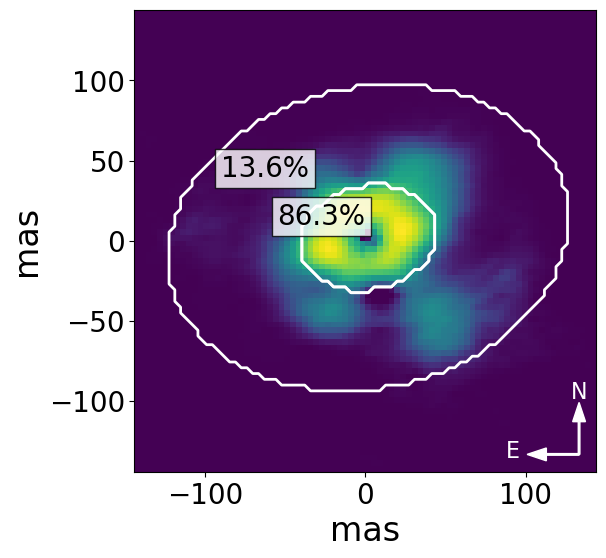}
    
    \caption{Resolved disc substructures in $V$ (left) and $I'$ (right) bands. The top row presents $Q_\phi$ images with white lines indicating the local AoLP for resolved disc structures. The length of each white line is proportional to the degree of polarization in its respective region. The bottom row displays the percentage of polarized disc intensity per resolved structure for IRAS\,08544-4431 (see Section~\ref{sec:substr}).  All images are presented on an inverse hyperbolic scale and oriented North up and East to the left. The low intensity of the central 5x5 pixel region of each $Q_\phi$ image is a reduction bias caused by over-subtracting of unresolved central polarization (Section~\ref{sec:unres_polarization}).}
    \label{fig:substr}
\end{figure}

Firstly, we focused on regions of statistical significance in the $Q_\phi$ polarimetric images using the calculated SNR (SNR $\geq5$, see Section~\ref{sec:snr}). Next, we examined the AoLP for the resolved substructures (see Section~\ref{sec:data_reduction}), which reflects the orientation of the polarization vector in each region. In the top row of Fig.~\ref{fig:substr} we show the local angles of linear polarization over the resolved structures in the deconvolved $Q_\phi$ images. The clear centrosymmetric distribution of polarization vectors suggests low residual noise in these areas, confirming that the polarimetric signal predominantly arises from the single scattering of stellar light on the disc surface. Finally, we computed the percentage of total polarized intensity per resolved substructure excluding the unresolved central polarization (as shown in the bottom row of Fig.~\ref{fig:substr}), which provided the brightness of the disc substructures relative to other disc parts. 

Based on the findings above, we found reliable substructures in polarimetric images of the IRAS\,08544-4431 circumbinary disc in both $V$ and $I'-$bands. Notably, the substructures are more pronounced in the $I'-$band, accounting for 13.6$\%$ of the polarized disc brightness, compared to only 2.5$\%$ in the $V-$band.

\subsection{Investigating wavelength dependency of disc polarimetry}
\label{sec:wavelength}

In this section, we investigate the wavelength dependency of the IRAS\,08544-4431 circumbinary disc polarisation, examining its magnitude and resolved morphology. 

\subsubsection{Measuring polarimetric disc colour}

Polarimetric observations provide an estimate of the amount of light polarized by the disc surface layers, thereby offering a lower limit for the reflected light from the disc surface \citep[e.g.,][]{Benisty2022arXiv220309991B}. However, accurately characterizing the dust in the disc via polarimetry, particularly for distant targets like post-AGB circumbinary discs, poses challenges due to factors such as small angular disc size, instrumental effects, and the not fully established disc geometry. Nevertheless, we can measure the total polarized disc brightness across different wavelengths. Assuming similar scattering geometries across these wavelengths, the observed differences in the polarized signal are primarily caused by the scattering and absorption properties of the dust in the disc. Therefore, the wavelength dependence of polarized disc brightness serves as an excellent indicator for investigating dust properties \citep[e.g.,][]{Ma2023A&A...676A...6M, Ma2024A&A...683A..18M}.

\begin{figure} 
    \includegraphics[width=1\columnwidth]{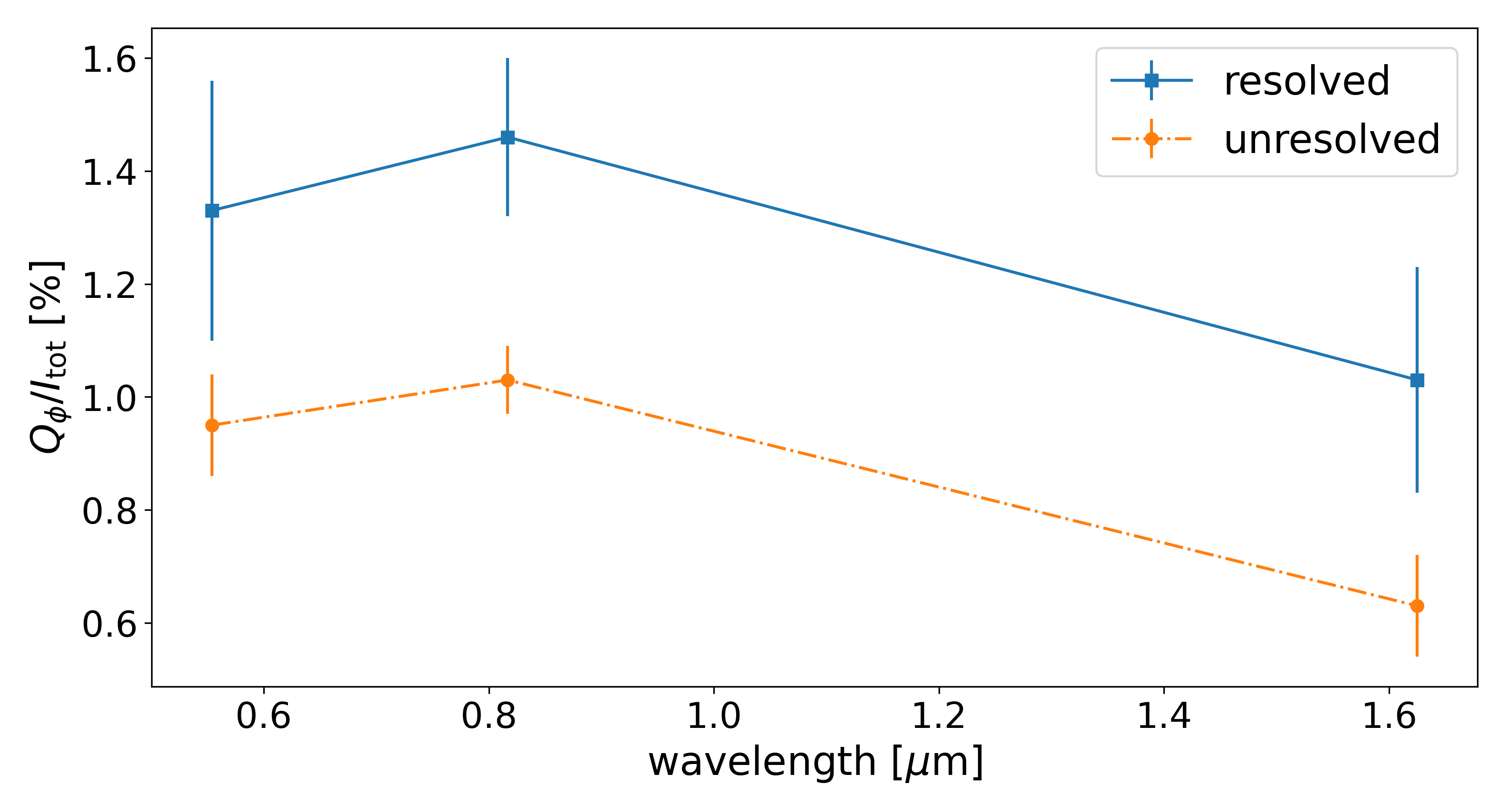}

    \caption{Resolved polarized disc brightness (solid blue line) and degree of unresolved polarisation (dashed orange line) relative to the total intensity of IRAS\,08544-4431 as a function of wavelength. See Section~\ref{sec:wavelength} for more details.}
    \label{fig:wavelength}
\end{figure}

To establish the wavelength dependence of the disc polarized intensity, we combined results from this study in the $V-$ and $I'-$bands, along with SPHERE/IRDIS $H-$band observations (see Fig.~\ref{fig:wavelength}). We note that the disc polarized brightness discussed in \citet{Andrych2023MNRAS.524.4168A} was not adjusted for the PSF smearing effect (see Section~\ref{sec:psf_smearing}). Therefore, we performed the additional correction of the $H$-band polarimetric results to ensure a consistent comparison with the disc polarized brightness in both the $V$ and $I'$ bands. To quantitatively characterize the wavelength dependence of the polarized disc brightness, we use the logarithmic wavelength gradient of the polarized reflectivity between two wavelengths as a measure of disc polarimetric colour:

\begin{equation}
    \eta_{\lambda_2/\lambda_1}=\frac{\rm log(Q_\phi/I_{tot})_{\lambda_1}-log(Q_\phi/I_{tot})_{\lambda_2}}{\rm log(\lambda_2/\lambda_1)},
	\label{eq:color}
\end{equation}

where $\lambda_1 < \lambda_2$, and $-0.5 < \eta < 0.5$ is classified as grey colour  \citep{Tazaki2019MNRAS.485.4951T, Ma2023A&A...676A...6M}.

 We found  that IRAS\,08544-4431 shows $\eta_{V/I}=0.24\pm0.72$ and $\eta_{V/H}=-0.24\pm0.34$ which is consistent with grey disc colour.

\subsubsection{Wavelength dependence of disc morphology}
\label{sec:col_index}

In addition to the wavelength dependence of the polarized disc brightness, we observed slight variations in the resolved disc morphology with wavelength. To investigate this effect, we initially compared the fitting parameters for the closest resolved section of the circumbinary disc to the central binary (bright 'ring'). We found that the resolved disc surface appears more elliptical towards lower wavelengths (see Fig.\ref{fig:disc_orient} and Table\ref{tab:disc_orient}), resulting in a higher estimation of the inclination of the resolved disc surface (see Section~\ref{sec:disc_orient}).

\begin{figure} 
    \includegraphics[width=1\columnwidth]{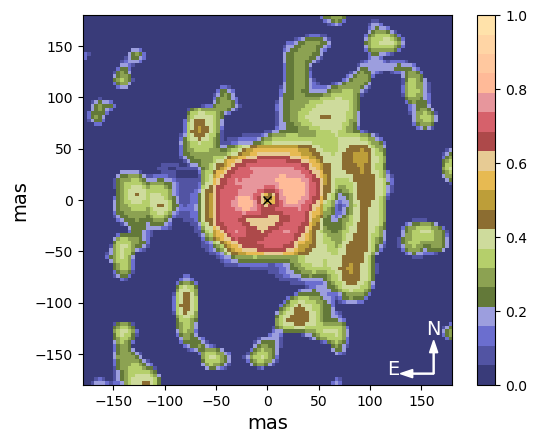}

    \caption{Combined $Q_\phi$ polarized disc morphology for IRAS\,08544-4431 using $V$ and $I-$band data from this study, along with the SPHERE/IRDIS $H-$band results (adapted from \citet{Andrych2023MNRAS.524.4168A}). The black cross represents the position of the binary star. Before combining the $Q_\phi$ images in each band, they were normalised to the total intensity, corrected for the separation-dependent drop-off in illumination by multiplying by $r^2$ and accounting for the system inclination defined for the resolved disc surface in each band (see Table.~\ref{tab:disc_orient}). Additionally, the image is presented on an inverse hyperbolic scale and normalised to highlight the intensity change along the disc. See Section~\ref{sec:wavelength} for more details.}
    \label{fig:wavelength_morph}
\end{figure}

Furthermore, we noticed that the polarimetric observations of the IRAS\,08544-4431 circumbinary disc show a more extended structure with longer wavelengths even though the polarized disc brightness is lower in near-IR. To investigate this, we combined $Q_\phi$ images of IRAS\,08544-4431 in $V$ and $I'-$band (from this study), along with the $Q_\phi$ image of the system in SPHERE/IRDIS $H-$band (adapted from \citet{Andrych2023MNRAS.524.4168A}). The $Q_\phi$ images were corrected for the separation-dependent drop-off in illumination by multiplying by $r^2$ and accounting for the system inclination defined for the resolved disc surface in each band (see Table~\ref{tab:disc_orient}). In Fig.~\ref{fig:wavelength_morph} we  present the final combined $Q_\phi$ images of IRAS\,08544-4431 in $V$, $I'$ and $H-$band. Additionally, we use an inverse hyperbolic scale and a discrete colour map to highlight the intensity change along the disc. 

To investigate the lack of extended structure in the $V-$ and $I'-$bands compared to previous $H-$band observations, we compared radial brightness profiles with the expected r$^{-2}$ illumination drop-off, typical for scattered light signals due to the dissipation of stellar illumination (see Fig.\ref{fig:profiles}). We found that in both $V$ and $I'-$bands, starting from $\sim 0.04"$, the radial brightness profiles drop off more steeply than expected. This suggests that the disc is either smaller than previously thought (which contradicts previous $H-$band observations), exhibiting a significant drop in the surface dust density or shadowed beyond the point where we detect polarized signal at a particular wavelength \citep[e.g.,][]{Perez2018ApJ...869L..50P}. We found the latter explanation to be more plausible, especially considering the lower opacity of the dust for longer wavelengths. We discuss possible causes of this effect in Section~\ref{sec:discussion}.

\section{Discussion}
\label{sec:discussion}

In this section, we interpret the obtained results to better understand the multi-wavelength morphology and dust scattering properties of the IRAS\,08544-4431 circumbinary disc surface. We achieve this by: i) analyzing the effect of interstellar polarization, ii) characterizing the dust properties on the disc surface, iii) investigating the variation in disc morphology with wavelength in SPHERE polarimetric imaging of IRAS\,08544-4431, and iv) exploring the similarities between the circumbinary disc around post-AGB binary stars and PPDs around YSOs.

\subsection{Effect of the interstellar polarisation}
\label{sec:interstellar}

As IRAS\,08544-4431 is located in the Galactic plane at distance of $\sim1600$ pc \citep{Bailer-Jones2021AJ....161..147B}, the observed polarized intensity of the disc can be significantly affected by scattering in the diffuse interstellar medium. The interstellar polarisation arises from differential extinction by aligned dust grains and depends on the degree of alignment with the local magnetic field, the angle between the local magnetic field and the line of sight, and the degree to which the direction of the magnetic field varies along the line of sight \citep{Draine2003ARA&A..41..241D}. Moreover, \citet{Ma2023A&A...676A...6M} showed that for the nearby PPD, such as RX J1604  ($\sim100$ pc), the unresolved polarisation is compatible with the interstellar polarization. 

During the data reduction process (see Section~\ref{sec:unres_polarization}), we computed the unresolved central polarization in our data and subtracted it from the final polarized images used for further analysis. Interstellar polarization has little to no impact on the integrated azimuthal polarization of the resolved disc surface, as it adds a butterfly-like pattern to $Q_\phi$ images with a net effect close to zero. However, this interstellar polarization is included in the measured unresolved central polarization. To determine whether this signal primarily originates from interstellar polarization or the unresolved disc component, we compared the unresolved central polarization with measurements of polarization from stars within a few degrees and roughly at the same distance as IRAS\,08544-4431 \citep[$\sim$1.5-2 kpc,][]{Bailer-Jones2021AJ....161..147B}. Galactic interstellar polarization surveys \citep{Heiles2000AJ....119..923H, Versteeg2023AJ....165...87V} indicate that stars within this range exhibit a degree of polarization in the range of 0.6-1.8$\%$ in $V-$band and a polarization angle of $85-130^\circ$.

\begin{figure} 
    \includegraphics[width=1\columnwidth]{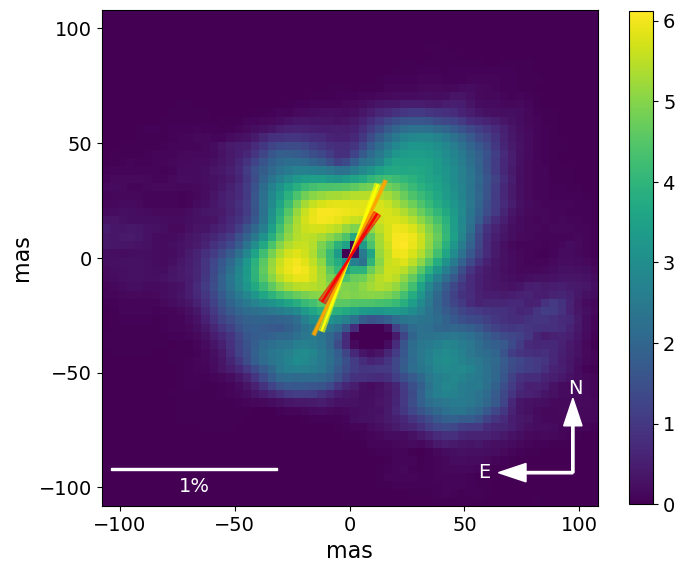}

    \caption{Orientation and relative degree of unresolved central polarisation of IRAS\,08544-4431 in $V-$, $I'-$, and $H-$bands presented over polarisation image of the system in SPHERE/ZIMPOL I-band. See Section~\ref{sec:interstellar} for more details.}
    \label{fig:unresolved}
\end{figure}

We found that the angle of unresolved central polarization for IRAS\,08544-4431 does not show significant rotation with wavelengths (see Fig.~\ref{fig:unresolved}). Although the measured unresolved central polarization for IRAS\,08544-4431 falls within the range of observed interstellar polarization in the region (see Table~\ref{tab:unresolved}), the calculated measurements do not align with the typical wavelength dependence expected for interstellar polarization \citep[e.g.,][]{Serkowski1975ApJ...196..261S}. Moreover, we found that the wavelength dependence of resolved polarized disc brightness and unresolved central polarization for IRAS\,08544-4431 is very similar, providing further evidence for a common origin of the signal. Therefore, we conclude that the unresolved central polarization measured for IRAS\,08544-4431 is predominantly caused by the unresolved part of the circumbinary disc rather than polarization in the diffuse interstellar medium along the line of sight.

\subsection{Characterizing the dust properties on the disc surface}
\label{sec:dust}

The intensity and polarization degree of reflected stellar light, along with the angular dependence of these measures (phase functions), and their wavelength dependency, encode details about the properties of the dust within the circumstellar disc \citep[e.g.,][]{Benisty2022arXiv220309991B}.

Based on SPHERE polarimetric imaging observations of IRAS\,08544-4431, we observed substantial polarization (up to 1.5\%) of the resolved disc surface in the optical and near-IR wavelengths. We also note that these values represent only lower limits due to the subtraction of unresolved central polarization (see Section~\ref{sec:unres_polarization}). This places the IRAS\,08544-4431 circumbinary disc among the brightest protoplanetary discs in terms of polarized intensity, with typical polarized intensity across the entire disc being approximately $10^{-2}$ of the stellar intensity \citep[e.g.,][]{Fukagawa2010PASJ...62..347F, Avenhaus2018ApJ...863...44A}. Although direct measurement of the polarization efficiency of the IRAS\,08544-4431 circumbinary disc is challenging due to the domination of the stellar PSF in the SPHERE total intensity image and the lack of resolved scattered light disc intensity (see Appendix~\ref{sec:scattered}), we can assume that its polarization efficiency is not less than the typical value for protoplanetary discs. Additionally, we found a clear brightness asymmetry in the forward and backward polarized scattering intensity in $V$ and $I'-$bands (see Section~\ref{sec:profiles}). This asymmetry suggests the presence of dust grains or aggregates with sizes at least in the micron range, exhibiting anisotropic scattering, contrary to isotropic Rayleigh scattering observed for dust grains smaller than the observing wavelength \citep[e.g.,][]{Ginski2023ApJ...953...92G}. Furthermore, the IRAS\,08544-4431 circumbinary disc appeared to have a grey colour in optical and near-IR polarimetry.

Recent theoretical studies on the relation between disc scattered-light (total intensity, polarized intensity, and colours) and dust properties \citep[size and structure; e.g.,][]{Mulders2013A&A...549A.112M, Tazaki2019MNRAS.485.4951T, Tazaki2022A&A...663A..57T} have shown that significant forward scattering, high polarization efficiency, and grey colour of the disc in optical and near-IR polarimetry can be best explained by large porous aggregates composed of submicron-sized monomers.

In conclusion, optical and near-IR SPHERE polarimetric observations of IRAS\,08544-4431 are compatible with the prevalence of large porous aggregates composed of submicron-sized monomers in the surface layers of the circumbinary disc. This characterization of dust scattering parameters paves the way for comparative analyses with protoplanetary disc systems around young stars and provides insights into dust evolution and disc dynamics.

\subsection{Investigating the complex disc morphology of IRAS\,08544-4431}
\label{sec:discus_morph}

Optical ($V$- and $I'$-band) polarimetric imaging of the full disc post-AGB system IRAS\,08544-4431 reveals a bright 'ring' with additional substructures (see Fig.~\ref{fig:disc_orient} and Section~\ref{sec:disc_orient}), which aligns well with our previous study using $H-$band polarimetric observations \citep{Andrych2023MNRAS.524.4168A}.
However, our analysis reveals that the 'ring' appears more eccentric towards shorter wavelengths, resulting in a higher estimated disc surface inclination (see Section~\ref{sec:morphology}). Furthermore, we noted a difference in the extended structure of the disc between various wavelength bands, with the near-IR $H-$band observations revealing a more extended structure compared to the optical $V$ and $I'-$bands (see Fig.\ref{fig:wavelength_morph}). As we discussed in Section~\ref{sec:dust}, polarimetric observations of IRAS\,08544-4431 are compatible with the prevalence of large porous dust aggregates composed of submicron-sized monomers in the surface layers of the circumbinary disc. Studies of the optical properties of porous dust aggregates reveal that dust opacity decreases from optical to near-IR wavelengths. This decrease potentially allows SPHERE near-IR polarimetric imaging to access slightly deeper disc layers than those accessible in the optical or to probe regions that were 'shadowed' in optical observations by low dust density clumps above the disc midplane \citep{Kirchschlager2014A&A...568A.103K, Tazaki2018ApJ...860...79T}. The above arguments lead us to speculate that the wavelength-dependant morphological differences observed for the IRAS\,08544-4431 disc could be explained by slightly misaligned or warped disc segments obscuring the extended disc structure at the shorter wavelengths (see Fig.~\ref{fig:substr}). Recent observational studies of 
circumstellar discs around YSOs \citep[e.g., ][]{Perez2018ApJ...869L..50P, Benisty2022arXiv220309991B} 
have revealed that disc warping is a common phenomenon. Warped segments of the disc can lead to shadows on the disc surface observed in scattered light. Numerical simulations suggested that disc warping may arise from gravitational interactions with an inclined binary system \citep{Young2022MNRAS.513..487Y,Young2023MNRAS.525.2616Y}, or from the influence of an inclined planet altering the disc structure \citep{Nealon2019MNRAS.484.4951N}. However, additional high-resolution observations of dust continuum and gas emission are required to further explore the midplane of the IRAS\,08544-4431 circumbinary disc and to better constrain the underlying disc structure and associated physical mechanisms. In our future study, we plan to use ALMA molecular line observations to investigate the velocity map structure, which could provide additional evidence of warping.

\subsection{Comparison of discs around post-AGBs with protoplanetary discs around YSOs}
\label{sec:comparison}

Advances in high-contrast imaging over the last decade have facilitated systematic observational studies of PPDs around YSOs, revealing a wide range of brightness, flux distribution, and disc extent \citep[e.g.,][]{Avenhaus2018ApJ...863...44A, Garufi2020A&A...633A..82G, Ginski2021ApJ...908L..25G}. Moreover, they have led to more precise modelling results, providing insights into dust properties \citep{Tazaki2019MNRAS.485.4951T, Tazaki2022A&A...663A..57T}. As mentioned in Section.~\ref{sec:intro}, post-AGB circumbinary discs appear remarkably similar to protoplanetary discs in terms of Keplerian rotation, IR excesses in the SED, disc dust mass, chemical depletion, dust mineralogy, and grain evolution, despite significant differences in formation history and lifetime ($\sim10^5$ years for post-AGB discs compared to a few million years for protoplanetary discs). 

Pilot high-angular resolution and polarimetric studies of post-AGB discs have revealed complex morphologies with arcs, gaps, and cavities, similar to those observed in protoplanetary discs \citep{Ertel2019AJ....157..110E, Andrych2023MNRAS.524.4168A}. Moreover, it has been found that post-AGB discs exhibit a similar level of near-IR polarized brightness relative to the stellar intensity ($\sim10^{-2}$) as previously shown for protoplanetary discs \citep[e.g.,][]{Avenhaus2018ApJ...863...44A, Ma2024A&A...683A..18M}. Using SPHERE/ZIMPOL polarimetric imaging data, we found that IRAS\,08544-4431 demonstrates a similar order of values for polarized brightness of the disc in optical wavelengths as well ($V$ and $I'-$bands, see Section.\ref{sec: polarized_bright}). Furthermore, employing the methodology presented by \citet{Garufi2014A&A...568A..40G, Garufi2017A&A...603A..21G}, we estimated the polarized contrast of the IRAS\,08544-4431 disc to be $\sim22\times 10^{-3}$ in the optical, positioning this system among the brightest protoplanetary discs \citep[see Section~\ref{sec: polarized_bright};][]{Garufi2022A&A...658A.137G, Benisty2022arXiv220309991B, Ginski2024arXiv240302149G}. However, it is important to note that the polarized contrast values presented in these studies should be treated as lower limits, as the reduction process did not include correction for some instrumental effects such as PSF smearing effects (see Section.\ref{sec:psf_smearing}).

\begin{figure} 
    \includegraphics[width=1\columnwidth]{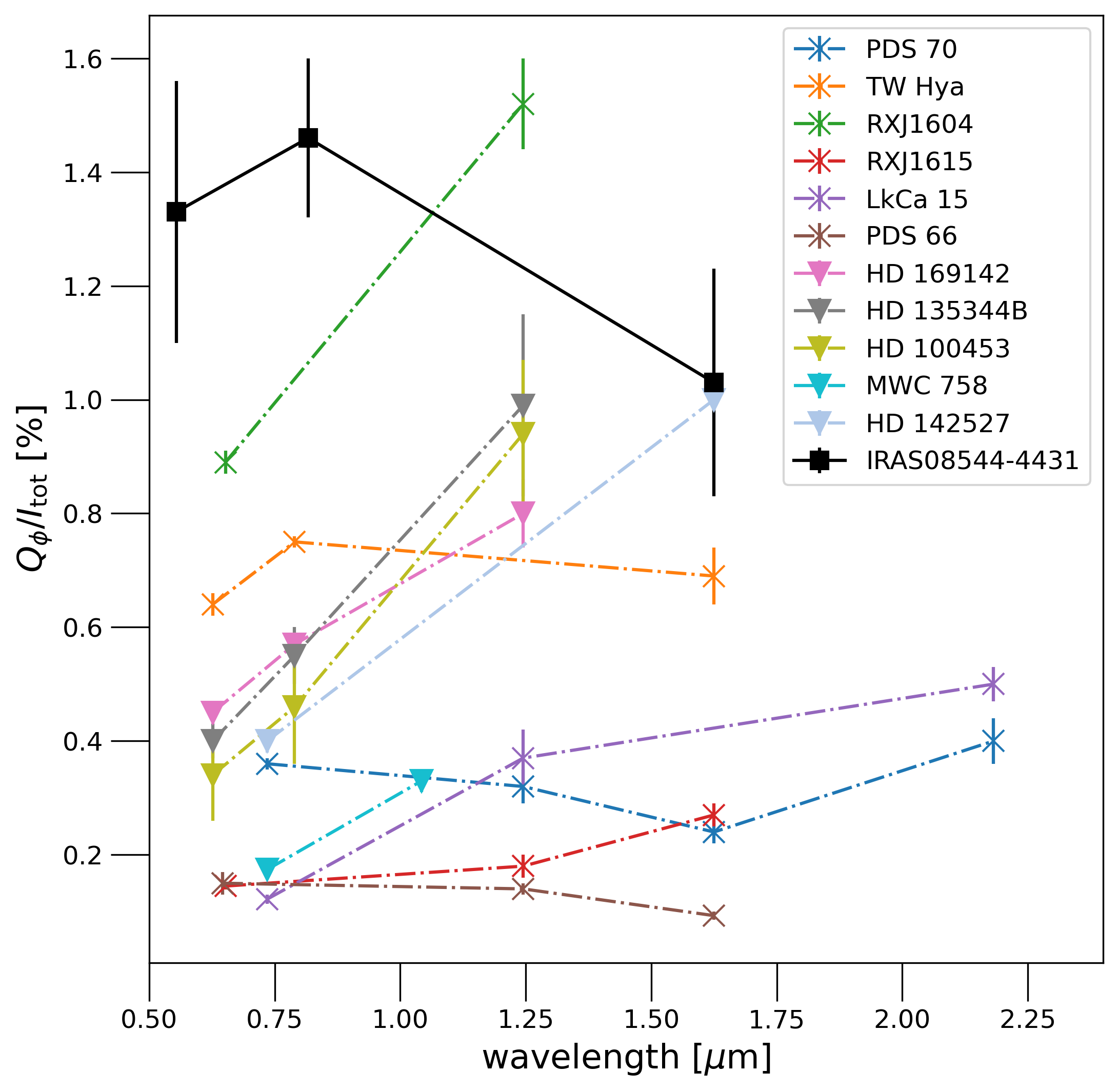}

    \caption{Polarized brightness measurements of the disc relative to the stellar intensity for IRAS\,08544-4431 (in black, this study) and a sample of young stellar objects \citep[T Tauri star discs marked with x and Herbig star discs with $\triangledown$][]{Ma2024A&A...683A..18M}. For a given disc the results for different wavelength bands are connected with lines. See Section~\ref{sec:comparison} for more details.}
    \label{fig:comparison}
\end{figure}

To compare our results for IRAS\,08544-4431 with existing accurate multi-wavelength measurements of PPD, we refer to the results of \citet{Ma2024A&A...683A..18M} for a sample of T Tauri and Herbig stars. In their study, authors focused on precise quantitative polarimetric analysis and considered the same instrumental effects as we do in this paper, ensuring a consistent comparison for the disc polarized brightness (see Section~\ref{sec:data_reduction}). In Fig.~\ref{fig:comparison}, we present the polarized brightness of the post-AGB system IRAS\,08544-4431  alongside a sample of Herbig and T Tauri stars. We note that the results for IRAS\,08544-4431 align well with those of the brightest protoplanetary discs. Furthermore, we observe a significant similarity in the wavelength dependence of disc polarization between IRAS\,08544-4431 and the T Tauri system TW\,Hya. 

TW\,Hya is a young stellar object surrounded by a gas-rich face-on protoplanetary disc \citep{Bergin2013Natur.493..644B, Macias2021A&A...648A..33M}. High angular resolution studies of the disc in thermal emission \citep{Nomura2016ApJ...819L...7N, Macias2021A&A...648A..33M} and scattered light \citep{vanBoekel2017ApJ...837..132V} have revealed multiple gaps in the disc. Additionally, observations of scattered light have unveiled an azimuthal variation in disc brightness caused by shadows cast by optically thick disc structures such as warps or misaligned rings \citep{Debes2023ApJ...948...36D, Teague2022ApJ...930..144T}. While direct confirmation of existing planets in the TW\,Hya disc is pending, recent observational and modelling studies suggest that the ring substructures within the circumstellar disc may serve as ideal sites to trigger the streaming instability and form new generations of planetesimals \citep[e.g.,][]{Mentiplay2019MNRAS.484L.130M, Macias2021A&A...648A..33M}. 

We note that both IRAS\,08544-4431 and TW\,Hya systems exhibit similar grey colours in optical and near-IR disc polarimetry. Moreover, $I'-$band (this study) and $H-$band \citep{Andrych2023MNRAS.524.4168A} polarimetric images revealed the asymmetric extended disc structure of IRAS08544-4431, suggesting further similarities. Although the distance to IRAS\,08544-4431 \citep[$\sim1.5$ kpc,][]{Bailer-Jones2021AJ....161..147B} prohibits us from studying the system at the same detailed level as TW\,Hya, the noted similarities provide further evidence that post-AGB circumbinary discs may be more similar to protoplanetary discs than previously thought. 

\section{Conclusions}
\label{sec:conclusion}

We present a multi-wavelength polarimetric imaging study of the post-AGB binary system IRAS\,08544-4431 using the VLT/SPHERE instrument. The aim of our study was threefold: to investigate the polarimetric properties and surface morphology of the disc across different wavelengths, to explore the corresponding dust properties encoded in both the intensity and degree of polarization of the scattered light and to compare our findings with those of protoplanetary discs. 

We successfully resolved the extended disc structure of the IRAS\,08544-4431 circumbinary disc in optical $V$ and $I'-$bands, confirming the complex disc morphology previously observed in near-IR polarimetric study. Despite the small angular size typical of post-AGB discs, IRAS\,08544-4431 exhibits significant brightness in azimuthal polarization at optical wavelengths, reaching up to $1.5\cdot10^{-2}$ (or $1.5\%$) of the total system intensity. Moreover, polarized observations of IRAS\,08544-4431 reveal significant forward scattering in optical polarimetry, suggesting that the northern part of the disc is closer to the observer. Additionally, we found that the disc shows a grey polarimetric colour in both optical and near-IR. These findings are consistent with theoretical models suggesting large porous aggregates of submicron-sized monomers as the dominant dust composition. Moreover, we found a variation in the extended structure of the disc across different wavelengths, with the near-IR $H-$band observations revealing a more extended disc surface compared to the optical $V$ and $I'-$bands.

Our study revealed similarities between the post-AGB circumbinary disc of IRAS\,08544-4431 and PPDs around YSOs. These similarities include comparable levels of disc polarized brightness, polarimetric colour, dust grain size and composition. We also suggest that a variation in the extended structure of the disc across different wavelengths could result from potential disc warping, similar to what has been proposed for certain protoplanetary discs. However, additional high-resolution observations of dust continuum and gas emission are required to further explore the mid-plane of the IRAS\,08544-4431 circumbinary disc and better constrain the underlying disc structure and associated physical mechanisms. 

In our following study, we will use both cameras of the SPHERE instrument to expand our multi-wavelength analysis to a larger sample of post-AGB binary systems. We aim to systematically investigate the wavelength-dependent dust scattering and polarization properties, multi-wavelength disc morphology, and estimate the dominant dust grain size on the circumbinary disc surface. This polarimetric study with SPHERE is part of a much larger observational campaign that includes optical, near-IR, and sub-mm interferometric observations, as well as spectral monitoring. This approach will enable us to better characterize the second-generation circumbinary discs around post-AGB binary stars, which is essential for studying disc dynamics across different stages of stellar evolution. Additionally, it offers a novel opportunity to investigate the potential formation of second-generation planets within dusty discs surrounding evolved stars.

\section*{Acknowledgements}

DK and KA acknowledge the support from the Australian Research Council Discovery Project DP240101150. This research was supported in part by the Australian Research Council Centre of Excellence for All Sky Astrophysics in 3 Dimensions (ASTRO 3D) through project number CE170100013. J.K. acknowledges support from FWO under the senior postdoctoral fellowship (1281121N). This study is based on observations collected at the European Southern Observatory under ESO programmes 101.D-0752(A), 0102.D-0696(A), and 0101.D-0807(B). This work has made use of the High Contrast Data Centre, jointly operated by OSUG/IPAG (Grenoble), PYTHEAS/LAM/CeSAM (Marseille), OCA/Lagrange (Nice), Observatoire de Paris/LESIA (Paris), and Observatoire de Lyon/CRAL, and supported by a grant from Labex OSUG@2020 (Investissements d’avenir – ANR10 LABX56). 

\section*{Data Availability}
The data underlying this article are stored online in the ESO Science Archive Facility at http://archive.eso.org, and can be accessed by program IDs.



\bibliographystyle{mnras}
\bibliography{andrych.bib} 

\begin{thebibliography}{}
\makeatletter
\relax
\def\mn@urlcharsother{\let\do\@makeother \do\$\do\&\do\#\do\^\do\_\do\%\do\~}
\def\mn@doi{\begingroup\mn@urlcharsother \@ifnextchar [ {\mn@doi@} {\mn@doi@[]}}
\def\mn@doi@[#1]#2{\def\@tempa{#1}\ifx\@tempa\@empty \href {http://dx.doi.org/#2} {doi:#2}\else \href {http://dx.doi.org/#2} {#1}\fi \endgroup}
\def\mn@eprint#1#2{\mn@eprint@#1:#2::\@nil}
\def\mn@eprint@arXiv#1{\href {http://arxiv.org/abs/#1} {{\tt arXiv:#1}}}
\def\mn@eprint@dblp#1{\href {http://dblp.uni-trier.de/rec/bibtex/#1.xml} {dblp:#1}}
\def\mn@eprint@#1:#2:#3:#4\@nil{\def\@tempa {#1}\def\@tempb {#2}\def\@tempc {#3}\ifx \@tempc \@empty \let \@tempc \@tempb \let \@tempb \@tempa \fi \ifx \@tempb \@empty \def\@tempb {arXiv}\fi \@ifundefined {mn@eprint@\@tempb}{\@tempb:\@tempc}{\expandafter \expandafter \csname mn@eprint@\@tempb\endcsname \expandafter{\@tempc}}}

\bibitem[\protect\citeauthoryear{{Andrych}, {Kamath}, {Kluska}, {Van Winckel}, {Ertel}  \& {Corporaal}}{{Andrych} et~al.}{2023}]{Andrych2023MNRAS.524.4168A}
{Andrych} K.,  {Kamath} D.,  {Kluska} J.,  {Van Winckel} H.,  {Ertel} S.,   {Corporaal} A.,  2023, \mn@doi [\mnras] {10.1093/mnras/stad1968}, \href {https://ui.adsabs.harvard.edu/abs/2023MNRAS.524.4168A} {524, 4168}

\bibitem[\protect\citeauthoryear{{Avenhaus} et~al.,}{{Avenhaus} et~al.}{2018}]{Avenhaus2018ApJ...863...44A}
{Avenhaus} H.,  et~al., 2018, \mn@doi [\apj] {10.3847/1538-4357/aab846}, \href {https://ui.adsabs.harvard.edu/abs/2018ApJ...863...44A} {863, 44}

\bibitem[\protect\citeauthoryear{{Bailer-Jones}, {Rybizki}, {Fouesneau}, {Demleitner}  \& {Andrae}}{{Bailer-Jones} et~al.}{2021}]{Bailer-Jones2021AJ....161..147B}
{Bailer-Jones} C.~A.~L.,  {Rybizki} J.,  {Fouesneau} M.,  {Demleitner} M.,   {Andrae} R.,  2021, \mn@doi [\aj] {10.3847/1538-3881/abd806}, \href {https://ui.adsabs.harvard.edu/abs/2021AJ....161..147B} {161, 147}

\bibitem[\protect\citeauthoryear{{Benisty} et~al.,}{{Benisty} et~al.}{2022}]{Benisty2022arXiv220309991B}
{Benisty} M.,  et~al., 2022, arXiv e-prints, \href {https://ui.adsabs.harvard.edu/abs/2022arXiv220309991B} {p. arXiv:2203.09991}

\bibitem[\protect\citeauthoryear{{Bergin} et~al.,}{{Bergin} et~al.}{2013}]{Bergin2013Natur.493..644B}
{Bergin} E.~A.,  et~al., 2013, \mn@doi [\nat] {10.1038/nature11805}, \href {https://ui.adsabs.harvard.edu/abs/2013Natur.493..644B} {493, 644}

\bibitem[\protect\citeauthoryear{{Beuzit} et~al.,}{{Beuzit} et~al.}{2019}]{Beuzit2019}
{Beuzit} J.~L.,  et~al., 2019, \mn@doi [\aap] {10.1051/0004-6361/201935251}, \href {https://ui.adsabs.harvard.edu/abs/2019A&A...631A.155B} {631, A155}

\bibitem[\protect\citeauthoryear{{Bollen}, {Kamath}, {Van Winckel}, {De Marco}, {Verhamme}, {Kluska}  \& {Wardle}}{{Bollen} et~al.}{2022}]{Bollen2022arXiv220808752B}
{Bollen} D.,  {Kamath} D.,  {Van Winckel} H.,  {De Marco} O.,  {Verhamme} O.,  {Kluska} J.,   {Wardle} M.,  2022, \mn@doi [\aap] {10.1051/0004-6361/202243429}, \href {https://ui.adsabs.harvard.edu/abs/2022A&A...666A..40B} {666, A40}

\bibitem[\protect\citeauthoryear{{Booth} \& {Owen}}{{Booth} \& {Owen}}{2020}]{Booth2020MNRAS.493.5079B}
{Booth} R.~A.,  {Owen} J.~E.,  2020, \mn@doi [\mnras] {10.1093/mnras/staa578}, \href {https://ui.adsabs.harvard.edu/abs/2020MNRAS.493.5079B} {493, 5079}

\bibitem[\protect\citeauthoryear{{Bujarrabal}, {Castro-Carrizo}, {Alcolea}, {Van Winckel}, {S{\'a}nchez Contreras}  \& {Santander-Garc{\'\i}a}}{{Bujarrabal} et~al.}{2017}]{Bujarrabal2017A&A...597L...5B}
{Bujarrabal} V.,  {Castro-Carrizo} A.,  {Alcolea} J.,  {Van Winckel} H.,  {S{\'a}nchez Contreras} C.,   {Santander-Garc{\'\i}a} M.,  2017, \mn@doi [\aap] {10.1051/0004-6361/201629317}, \href {https://ui.adsabs.harvard.edu/abs/2017A&A...597L...5B} {597, L5}

\bibitem[\protect\citeauthoryear{{Bujarrabal}, {Castro-Carrizo}, {Van Winckel}, {Alcolea}, {S{\'a}nchez Contreras}, {Santander-Garc{\'\i}a}  \& {Hillen}}{{Bujarrabal} et~al.}{2018}]{Bujarrabal2018}
{Bujarrabal} V.,  {Castro-Carrizo} A.,  {Van Winckel} H.,  {Alcolea} J.,  {S{\'a}nchez Contreras} C.,  {Santander-Garc{\'\i}a} M.,   {Hillen} M.,  2018, \mn@doi [\aap] {10.1051/0004-6361/201732422}, \href {https://ui.adsabs.harvard.edu/abs/2018A&A...614A..58B} {614, A58}

\bibitem[\protect\citeauthoryear{{Canovas}, {M{\'e}nard}, {de Boer}, {Pinte}, {Avenhaus}  \& {Schreiber}}{{Canovas} et~al.}{2015}]{Canovas2015A&A...582L...7C}
{Canovas} H.,  {M{\'e}nard} F.,  {de Boer} J.,  {Pinte} C.,  {Avenhaus} H.,   {Schreiber} M.~R.,  2015, \mn@doi [\aap] {10.1051/0004-6361/201527267}, \href {https://ui.adsabs.harvard.edu/abs/2015A&A...582L...7C} {582, L7}

\bibitem[\protect\citeauthoryear{{Corporaal}, {Kluska}, {Van Winckel}, {Bollen}, {Kamath}  \& {Min}}{{Corporaal} et~al.}{2021}]{Corporaal2021A&A...650L..13C}
{Corporaal} A.,  {Kluska} J.,  {Van Winckel} H.,  {Bollen} D.,  {Kamath} D.,   {Min} M.,  2021, \mn@doi [\aap] {10.1051/0004-6361/202141154}, \href {https://ui.adsabs.harvard.edu/abs/2021A&A...650L..13C} {650, L13}

\bibitem[\protect\citeauthoryear{{Corporaal}, {Kluska}, {Van Winckel}, {Kamath}  \& {Min}}{{Corporaal} et~al.}{2023a}]{Corporaal_IRAS08_2023A&A...671A..15C}
{Corporaal} A.,  {Kluska} J.,  {Van Winckel} H.,  {Kamath} D.,   {Min} M.,  2023a, \mn@doi [\aap] {10.1051/0004-6361/202245689}, \href {https://ui.adsabs.harvard.edu/abs/2023A&A...671A..15C} {671, A15}

\bibitem[\protect\citeauthoryear{{Corporaal}, {Kluska}, {Van Winckel}, {Andrych}, {Cuello}, {Kamath}  \& {M{\'e}rand}}{{Corporaal} et~al.}{2023b}]{Corporaal2023A&A...674A.151C}
{Corporaal} A.,  {Kluska} J.,  {Van Winckel} H.,  {Andrych} K.,  {Cuello} N.,  {Kamath} D.,   {M{\'e}rand} A.,  2023b, \mn@doi [\aap] {10.1051/0004-6361/202346408}, \href {https://ui.adsabs.harvard.edu/abs/2023A&A...674A.151C} {674, A151}

\bibitem[\protect\citeauthoryear{{De Prins}, {Van Winckel}, {Ferreira}, {Verhamme}, {Kamath}, {Zimniak}  \& {Jacquemin-Ide}}{{De Prins} et~al.}{2024}]{DePrins2024arXiv240609280D}
{De Prins} T.,  {Van Winckel} H.,  {Ferreira} J.,  {Verhamme} O.,  {Kamath} D.,  {Zimniak} N.,   {Jacquemin-Ide} J.,  2024, \mn@doi [arXiv e-prints] {10.48550/arXiv.2406.09280}, \href {https://ui.adsabs.harvard.edu/abs/2024arXiv240609280D} {p. arXiv:2406.09280}

\bibitem[\protect\citeauthoryear{{Debes} et~al.,}{{Debes} et~al.}{2023}]{Debes2023ApJ...948...36D}
{Debes} J.,  et~al., 2023, \mn@doi [\apj] {10.3847/1538-4357/acbdf1}, \href {https://ui.adsabs.harvard.edu/abs/2023ApJ...948...36D} {948, 36}

\bibitem[\protect\citeauthoryear{{Delorme} et~al.,}{{Delorme} et~al.}{2017}]{Delorme2017sf2a.conf..347D}
{Delorme} P.,  et~al., 2017, in {Reyl{\'e}} C.,  {Di Matteo} P.,  {Herpin} F.,  {Lagadec} E.,  {Lan{\c{c}}on} A.,  {Meliani} Z.,   {Royer} F.,  eds, SF2A-2017: Proceedings of the Annual meeting of the French Society of Astronomy and Astrophysics. p.~Di (\mn@eprint {arXiv} {1712.06948}), \mn@doi{10.48550/arXiv.1712.06948}

\bibitem[\protect\citeauthoryear{{Dohlen} et~al.,}{{Dohlen} et~al.}{2008}]{Dohlen2008}
{Dohlen} K.,  et~al., 2008, in {McLean} I.~S.,  {Casali} M.~M.,  eds,  Society of Photo-Optical Instrumentation Engineers (SPIE) Conference Series Vol. 7014, Ground-based and Airborne Instrumentation for Astronomy II. p. 70143L, \mn@doi{10.1117/12.789786}

\bibitem[\protect\citeauthoryear{{Draine}}{{Draine}}{2003}]{Draine2003ARA&A..41..241D}
{Draine} B.~T.,  2003, \mn@doi [\araa] {10.1146/annurev.astro.41.011802.094840}, \href {https://ui.adsabs.harvard.edu/abs/2003ARA&A..41..241D} {41, 241}

\bibitem[\protect\citeauthoryear{{Ertel} et~al.,}{{Ertel} et~al.}{2019}]{Ertel2019AJ....157..110E}
{Ertel} S.,  et~al., 2019, \mn@doi [\aj] {10.3847/1538-3881/aafe04}, \href {https://ui.adsabs.harvard.edu/abs/2019AJ....157..110E} {157, 110}

\bibitem[\protect\citeauthoryear{{Fukagawa} et~al.,}{{Fukagawa} et~al.}{2010}]{Fukagawa2010PASJ...62..347F}
{Fukagawa} M.,  et~al., 2010, \mn@doi [\pasj] {10.1093/pasj/62.2.347}, \href {https://ui.adsabs.harvard.edu/abs/2010PASJ...62..347F} {62, 347}

\bibitem[\protect\citeauthoryear{{Gallardo Cava}, {G{\'o}mez-Garrido}, {Bujarrabal}, {Castro-Carrizo}, {Alcolea}  \& {Van Winckel}}{{Gallardo Cava} et~al.}{2021}]{Gallardo_cava2021A&A...648A..93G}
{Gallardo Cava} I.,  {G{\'o}mez-Garrido} M.,  {Bujarrabal} V.,  {Castro-Carrizo} A.,  {Alcolea} J.,   {Van Winckel} H.,  2021, \mn@doi [\aap] {10.1051/0004-6361/202039604}, \href {https://ui.adsabs.harvard.edu/abs/2021A&A...648A..93G} {648, A93}

\bibitem[\protect\citeauthoryear{{Garufi}, {Quanz}, {Schmid}, {Avenhaus}, {Buenzli}  \& {Wolf}}{{Garufi} et~al.}{2014}]{Garufi2014A&A...568A..40G}
{Garufi} A.,  {Quanz} S.~P.,  {Schmid} H.~M.,  {Avenhaus} H.,  {Buenzli} E.,   {Wolf} S.,  2014, \mn@doi [\aap] {10.1051/0004-6361/201424262}, \href {https://ui.adsabs.harvard.edu/abs/2014A&A...568A..40G} {568, A40}

\bibitem[\protect\citeauthoryear{{Garufi} et~al.,}{{Garufi} et~al.}{2017}]{Garufi2017A&A...603A..21G}
{Garufi} A.,  et~al., 2017, \mn@doi [\aap] {10.1051/0004-6361/201630320}, \href {https://ui.adsabs.harvard.edu/abs/2017A&A...603A..21G} {603, A21}

\bibitem[\protect\citeauthoryear{{Garufi} et~al.,}{{Garufi} et~al.}{2020}]{Garufi2020A&A...633A..82G}
{Garufi} A.,  et~al., 2020, \mn@doi [\aap] {10.1051/0004-6361/201936946}, \href {https://ui.adsabs.harvard.edu/abs/2020A&A...633A..82G} {633, A82}

\bibitem[\protect\citeauthoryear{{Garufi} et~al.,}{{Garufi} et~al.}{2022}]{Garufi2022A&A...658A.137G}
{Garufi} A.,  et~al., 2022, \mn@doi [\aap] {10.1051/0004-6361/202141692}, \href {https://ui.adsabs.harvard.edu/abs/2022A&A...658A.137G} {658, A137}

\bibitem[\protect\citeauthoryear{{Gielen} et~al.,}{{Gielen} et~al.}{2011}]{Gielen2011A&A...533A..99G}
{Gielen} C.,  et~al., 2011, \mn@doi [\aap] {10.1051/0004-6361/201117364}, \href {https://ui.adsabs.harvard.edu/abs/2011A&A...533A..99G} {533, A99}

\bibitem[\protect\citeauthoryear{{Ginski} et~al.,}{{Ginski} et~al.}{2021}]{Ginski2021ApJ...908L..25G}
{Ginski} C.,  et~al., 2021, \mn@doi [\apjl] {10.3847/2041-8213/abdf57}, \href {https://ui.adsabs.harvard.edu/abs/2021ApJ...908L..25G} {908, L25}

\bibitem[\protect\citeauthoryear{{Ginski}, {Tazaki}, {Dominik}  \& {Stolker}}{{Ginski} et~al.}{2023}]{Ginski2023ApJ...953...92G}
{Ginski} C.,  {Tazaki} R.,  {Dominik} C.,   {Stolker} T.,  2023, \mn@doi [\apj] {10.3847/1538-4357/acdc97}, \href {https://ui.adsabs.harvard.edu/abs/2023ApJ...953...92G} {953, 92}

\bibitem[\protect\citeauthoryear{{Ginski} et~al.,}{{Ginski} et~al.}{2024}]{Ginski2024arXiv240302149G}
{Ginski} C.,  et~al., 2024, \mn@doi [arXiv e-prints] {10.48550/arXiv.2403.02149}, \href {https://ui.adsabs.harvard.edu/abs/2024arXiv240302149G} {p. arXiv:2403.02149}

\bibitem[\protect\citeauthoryear{{Hardy} et~al.,}{{Hardy} et~al.}{2016}]{Marsh2016MNRAS.459.4518H}
{Hardy} A.,  et~al., 2016, \mn@doi [\mnras] {10.1093/mnras/stw976}, \href {https://ui.adsabs.harvard.edu/abs/2016MNRAS.459.4518H} {459, 4518}

\bibitem[\protect\citeauthoryear{{Heiles}}{{Heiles}}{2000}]{Heiles2000AJ....119..923H}
{Heiles} C.,  2000, \mn@doi [\aj] {10.1086/301236}, \href {https://ui.adsabs.harvard.edu/abs/2000AJ....119..923H} {119, 923}

\bibitem[\protect\citeauthoryear{{Hillen}, {Kluska}, {Le Bouquin}, {Van Winckel}, {Berger}, {Kamath}  \& {Bujarrabal}}{{Hillen} et~al.}{2016}]{Hillen2016}
{Hillen} M.,  {Kluska} J.,  {Le Bouquin} J.~B.,  {Van Winckel} H.,  {Berger} J.~P.,  {Kamath} D.,   {Bujarrabal} V.,  2016, \mn@doi [\aap] {10.1051/0004-6361/201628125}, \href {https://ui.adsabs.harvard.edu/abs/2016A&A...588L...1H} {588, L1}

\bibitem[\protect\citeauthoryear{{Hunziker} et~al.,}{{Hunziker} et~al.}{2021}]{Hunziker2021A&A...648A.110H}
{Hunziker} S.,  et~al., 2021, \mn@doi [\aap] {10.1051/0004-6361/202040166}, \href {https://ui.adsabs.harvard.edu/abs/2021A&A...648A.110H} {648, A110}

\bibitem[\protect\citeauthoryear{{Kamath}, {Wood}  \& {Van Winckel}}{{Kamath} et~al.}{2015}]{Kamath2015MNRAS.454.1468K}
{Kamath} D.,  {Wood} P.~R.,   {Van Winckel} H.,  2015, \mn@doi [\mnras] {10.1093/mnras/stv1202}, \href {https://ui.adsabs.harvard.edu/abs/2015MNRAS.454.1468K} {454, 1468}

\bibitem[\protect\citeauthoryear{{Keppler} et~al.,}{{Keppler} et~al.}{2018}]{Keppler2018A&A...617A..44K}
{Keppler} M.,  et~al., 2018, \mn@doi [\aap] {10.1051/0004-6361/201832957}, \href {https://ui.adsabs.harvard.edu/abs/2018A&A...617A..44K} {617, A44}

\bibitem[\protect\citeauthoryear{{Kirchschlager} \& {Wolf}}{{Kirchschlager} \& {Wolf}}{2014}]{Kirchschlager2014A&A...568A.103K}
{Kirchschlager} F.,  {Wolf} S.,  2014, \mn@doi [\aap] {10.1051/0004-6361/201323176}, \href {https://ui.adsabs.harvard.edu/abs/2014A&A...568A.103K} {568, A103}

\bibitem[\protect\citeauthoryear{{Kluska}, {Hillen}, {Van Winckel}, {Manick}, {Min}, {Regibo}  \& {Royer}}{{Kluska} et~al.}{2018}]{Kluska2018A&A...616A.153K}
{Kluska} J.,  {Hillen} M.,  {Van Winckel} H.,  {Manick} R.,  {Min} M.,  {Regibo} S.,   {Royer} P.,  2018, \mn@doi [\aap] {10.1051/0004-6361/201832983}, \href {https://ui.adsabs.harvard.edu/abs/2018A&A...616A.153K} {616, A153}

\bibitem[\protect\citeauthoryear{{Kluska}, {Van Winckel}, {Hillen}, {Berger}, {Kamath}, {Le Bouquin}  \& {Min}}{{Kluska} et~al.}{2019}]{Kluska2019A&A...631A.108K}
{Kluska} J.,  {Van Winckel} H.,  {Hillen} M.,  {Berger} J.~P.,  {Kamath} D.,  {Le Bouquin} J.~B.,   {Min} M.,  2019, \mn@doi [\aap] {10.1051/0004-6361/201935785}, \href {https://ui.adsabs.harvard.edu/abs/2019A&A...631A.108K} {631, A108}

\bibitem[\protect\citeauthoryear{{Kluska}, {Van Winckel}, {Copp{\'e}e}, {Oomen}, {Dsilva}, {Kamath}, {Bujarrabal}  \& {Min}}{{Kluska} et~al.}{2022}]{Kluska2022}
{Kluska} J.,  {Van Winckel} H.,  {Copp{\'e}e} Q.,  {Oomen} G.~M.,  {Dsilva} K.,  {Kamath} D.,  {Bujarrabal} V.,   {Min} M.,  2022, \mn@doi [\aap] {10.1051/0004-6361/202141690}, \href {https://ui.adsabs.harvard.edu/abs/2022A&A...658A..36K} {658, A36}

\bibitem[\protect\citeauthoryear{{Lucy}}{{Lucy}}{1974}]{Lucy1974AJ.....79..745L}
{Lucy} L.~B.,  1974, \mn@doi [\aj] {10.1086/111605}, \href {https://ui.adsabs.harvard.edu/abs/1974AJ.....79..745L} {79, 745}

\bibitem[\protect\citeauthoryear{{Ma}, {Schmid}  \& {Tschudi}}{{Ma} et~al.}{2023}]{Ma2023A&A...676A...6M}
{Ma} J.,  {Schmid} H.~M.,   {Tschudi} C.,  2023, \mn@doi [\aap] {10.1051/0004-6361/202245697}, \href {https://ui.adsabs.harvard.edu/abs/2023A&A...676A...6M} {676, A6}

\bibitem[\protect\citeauthoryear{{Ma}, {Schmid}  \& {Stolker}}{{Ma} et~al.}{2024}]{Ma2024A&A...683A..18M}
{Ma} J.,  {Schmid} H.~M.,   {Stolker} T.,  2024, \mn@doi [\aap] {10.1051/0004-6361/202347782}, \href {https://ui.adsabs.harvard.edu/abs/2024A&A...683A..18M} {683, A18}

\bibitem[\protect\citeauthoryear{{Maas}, {Van Winckel}  \& {Lloyd Evans}}{{Maas} et~al.}{2005}]{Maas2005}
{Maas} T.,  {Van Winckel} H.,   {Lloyd Evans} T.,  2005, \mn@doi [\aap] {10.1051/0004-6361:20041688}, \href {https://ui.adsabs.harvard.edu/abs/2005A&A...429..297M} {429, 297}

\bibitem[\protect\citeauthoryear{{Maaskant} et~al.,}{{Maaskant} et~al.}{2013}]{Maaskant2013A&A...555A..64M}
{Maaskant} K.~M.,  et~al., 2013, \mn@doi [\aap] {10.1051/0004-6361/201321300}, \href {https://ui.adsabs.harvard.edu/abs/2013A&A...555A..64M} {555, A64}

\bibitem[\protect\citeauthoryear{{Mac{\'\i}as}, {Guerra-Alvarado}, {Carrasco-Gonz{\'a}lez}, {Ribas}, {Espaillat}, {Huang}  \& {Andrews}}{{Mac{\'\i}as} et~al.}{2021}]{Macias2021A&A...648A..33M}
{Mac{\'\i}as} E.,  {Guerra-Alvarado} O.,  {Carrasco-Gonz{\'a}lez} C.,  {Ribas} {\'A}.,  {Espaillat} C.~C.,  {Huang} J.,   {Andrews} S.~M.,  2021, \mn@doi [\aap] {10.1051/0004-6361/202039812}, \href {https://ui.adsabs.harvard.edu/abs/2021A&A...648A..33M} {648, A33}

\bibitem[\protect\citeauthoryear{{Mentiplay}, {Price}  \& {Pinte}}{{Mentiplay} et~al.}{2019}]{Mentiplay2019MNRAS.484L.130M}
{Mentiplay} D.,  {Price} D.~J.,   {Pinte} C.,  2019, \mn@doi [\mnras] {10.1093/mnrasl/sly209}, \href {https://ui.adsabs.harvard.edu/abs/2019MNRAS.484L.130M} {484, L130}

\bibitem[\protect\citeauthoryear{{Menu}, {van Boekel}, {Henning}, {Leinert}, {Waelkens}  \& {Waters}}{{Menu} et~al.}{2015}]{Menu2015A&A...581A.107M}
{Menu} J.,  {van Boekel} R.,  {Henning} T.,  {Leinert} C.,  {Waelkens} C.,   {Waters} L.~B.~F.~M.,  2015, \mn@doi [\aap] {10.1051/0004-6361/201525654}, \href {https://ui.adsabs.harvard.edu/abs/2015A&A...581A.107M} {581, A107}

\bibitem[\protect\citeauthoryear{{Miller Bertolami}}{{Miller Bertolami}}{2016}]{Bertolami2016A&A...588A..25M}
{Miller Bertolami} M.~M.,  2016, \mn@doi [\aap] {10.1051/0004-6361/201526577}, \href {https://ui.adsabs.harvard.edu/abs/2016A&A...588A..25M} {588, A25}

\bibitem[\protect\citeauthoryear{{Mohorian}, {Kamath}, {Menon}, {Ventura}, {Van Winckel}, {Garc{\'\i}a-Hern{\'a}ndez}  \& {Masseron}}{{Mohorian} et~al.}{2024}]{Mohorian2024MNRAS.530..761M}
{Mohorian} M.,  {Kamath} D.,  {Menon} M.,  {Ventura} P.,  {Van Winckel} H.,  {Garc{\'\i}a-Hern{\'a}ndez} D.~A.,   {Masseron} T.,  2024, \mn@doi [\mnras] {10.1093/mnras/stae791}, \href {https://ui.adsabs.harvard.edu/abs/2024MNRAS.530..761M} {530, 761}

\bibitem[\protect\citeauthoryear{{Mulders}, {Min}, {Dominik}, {Debes}  \& {Schneider}}{{Mulders} et~al.}{2013}]{Mulders2013A&A...549A.112M}
{Mulders} G.~D.,  {Min} M.,  {Dominik} C.,  {Debes} J.~H.,   {Schneider} G.,  2013, \mn@doi [\aap] {10.1051/0004-6361/201219522}, \href {https://ui.adsabs.harvard.edu/abs/2013A&A...549A.112M} {549, A112}

\bibitem[\protect\citeauthoryear{{Nealon}, {Pinte}, {Alexander}, {Mentiplay}  \& {Dipierro}}{{Nealon} et~al.}{2019}]{Nealon2019MNRAS.484.4951N}
{Nealon} R.,  {Pinte} C.,  {Alexander} R.,  {Mentiplay} D.,   {Dipierro} G.,  2019, \mn@doi [\mnras] {10.1093/mnras/stz346}, \href {https://ui.adsabs.harvard.edu/abs/2019MNRAS.484.4951N} {484, 4951}

\bibitem[\protect\citeauthoryear{{Nomura} et~al.,}{{Nomura} et~al.}{2016}]{Nomura2016ApJ...819L...7N}
{Nomura} H.,  et~al., 2016, \mn@doi [\apjl] {10.3847/2041-8205/819/1/L7}, \href {https://ui.adsabs.harvard.edu/abs/2016ApJ...819L...7N} {819, L7}

\bibitem[\protect\citeauthoryear{{Oomen}, {Van Winckel}, {Pols}, {Nelemans}, {Escorza}, {Manick}, {Kamath}  \& {Waelkens}}{{Oomen} et~al.}{2018}]{Oomen2018}
{Oomen} G.-M.,  {Van Winckel} H.,  {Pols} O.,  {Nelemans} G.,  {Escorza} A.,  {Manick} R.,  {Kamath} D.,   {Waelkens} C.,  2018, \mn@doi [\aap] {10.1051/0004-6361/201833816}, \href {https://ui.adsabs.harvard.edu/abs/2018A&A...620A..85O} {620, A85}

\bibitem[\protect\citeauthoryear{{Oomen}, {Van Winckel}, {Pols}  \& {Nelemans}}{{Oomen} et~al.}{2019}]{Oomen2019A&A...629A..49O}
{Oomen} G.-M.,  {Van Winckel} H.,  {Pols} O.,   {Nelemans} G.,  2019, \mn@doi [\aap] {10.1051/0004-6361/201935853}, \href {https://ui.adsabs.harvard.edu/abs/2019A&A...629A..49O} {629, A49}

\bibitem[\protect\citeauthoryear{{P{\'e}rez} et~al.,}{{P{\'e}rez} et~al.}{2018}]{Perez2018ApJ...869L..50P}
{P{\'e}rez} L.~M.,  et~al., 2018, \mn@doi [\apjl] {10.3847/2041-8213/aaf745}, \href {https://ui.adsabs.harvard.edu/abs/2018ApJ...869L..50P} {869, L50}

\bibitem[\protect\citeauthoryear{{Richardson}}{{Richardson}}{1972}]{Richardson1972JOSA...62...55R}
{Richardson} W.~H.,  1972, Journal of the Optical Society of America (1917-1983), \href {https://ui.adsabs.harvard.edu/abs/1972JOSA...62...55R} {62, 55}

\bibitem[\protect\citeauthoryear{{Schleicher}, {Dreizler}, {V{\"o}lschow}, {Banerjee}  \& {Hessman}}{{Schleicher} et~al.}{2015}]{Schleicher2015AN....336..458S}
{Schleicher} D.~R.~G.,  {Dreizler} S.,  {V{\"o}lschow} M.,  {Banerjee} R.,   {Hessman} F.~V.,  2015, \mn@doi [Astronomische Nachrichten] {10.1002/asna.201412184}, \href {https://ui.adsabs.harvard.edu/abs/2015AN....336..458S} {336, 458}

\bibitem[\protect\citeauthoryear{{Schmid}, {Joos}  \& {Tschan}}{{Schmid} et~al.}{2006}]{Schmid2006A&A...452..657S}
{Schmid} H.~M.,  {Joos} F.,   {Tschan} D.,  2006, \mn@doi [\aap] {10.1051/0004-6361:20053273}, \href {https://ui.adsabs.harvard.edu/abs/2006A&A...452..657S} {452, 657}

\bibitem[\protect\citeauthoryear{{Schmid} et~al.,}{{Schmid} et~al.}{2018}]{Schmid2018A&A...619A...9S}
{Schmid} H.~M.,  et~al., 2018, \mn@doi [\aap] {10.1051/0004-6361/201833620}, \href {https://ui.adsabs.harvard.edu/abs/2018A&A...619A...9S} {619, A9}

\bibitem[\protect\citeauthoryear{{Scicluna}, {Kemper}, {Trejo}, {Marshall}, {Ertel}  \& {Hillen}}{{Scicluna} et~al.}{2020}]{Scicluna2020MNRAS.494.2925S}
{Scicluna} P.,  {Kemper} F.,  {Trejo} A.,  {Marshall} J.~P.,  {Ertel} S.,   {Hillen} M.,  2020, \mn@doi [\mnras] {10.1093/mnras/staa425}, \href {https://ui.adsabs.harvard.edu/abs/2020MNRAS.494.2925S} {494, 2925}

\bibitem[\protect\citeauthoryear{{Serkowski}, {Mathewson}  \& {Ford}}{{Serkowski} et~al.}{1975}]{Serkowski1975ApJ...196..261S}
{Serkowski} K.,  {Mathewson} D.~S.,   {Ford} V.~L.,  1975, \mn@doi [\apj] {10.1086/153410}, \href {https://ui.adsabs.harvard.edu/abs/1975ApJ...196..261S} {196, 261}

\bibitem[\protect\citeauthoryear{{Simmons} \& {Stewart}}{{Simmons} \& {Stewart}}{1985}]{Simmons1985A&A...142..100S}
{Simmons} J.~F.~L.,  {Stewart} B.~G.,  1985, \aap, \href {https://ui.adsabs.harvard.edu/abs/1985A&A...142..100S} {142, 100}

\bibitem[\protect\citeauthoryear{{Tazaki} \& {Dominik}}{{Tazaki} \& {Dominik}}{2022}]{Tazaki2022A&A...663A..57T}
{Tazaki} R.,  {Dominik} C.,  2022, \mn@doi [\aap] {10.1051/0004-6361/202243485}, \href {https://ui.adsabs.harvard.edu/abs/2022A&A...663A..57T} {663, A57}

\bibitem[\protect\citeauthoryear{{Tazaki} \& {Tanaka}}{{Tazaki} \& {Tanaka}}{2018}]{Tazaki2018ApJ...860...79T}
{Tazaki} R.,  {Tanaka} H.,  2018, \mn@doi [\apj] {10.3847/1538-4357/aac32d}, \href {https://ui.adsabs.harvard.edu/abs/2018ApJ...860...79T} {860, 79}

\bibitem[\protect\citeauthoryear{{Tazaki}, {Tanaka}, {Muto}, {Kataoka}  \& {Okuzumi}}{{Tazaki} et~al.}{2019}]{Tazaki2019MNRAS.485.4951T}
{Tazaki} R.,  {Tanaka} H.,  {Muto} T.,  {Kataoka} A.,   {Okuzumi} S.,  2019, \mn@doi [\mnras] {10.1093/mnras/stz662}, \href {https://ui.adsabs.harvard.edu/abs/2019MNRAS.485.4951T} {485, 4951}

\bibitem[\protect\citeauthoryear{{Teague}, {Bae}, {Benisty}, {Andrews}, {Facchini}, {Huang}  \& {Wilner}}{{Teague} et~al.}{2022}]{Teague2022ApJ...930..144T}
{Teague} R.,  {Bae} J.,  {Benisty} M.,  {Andrews} S.~M.,  {Facchini} S.,  {Huang} J.,   {Wilner} D.,  2022, \mn@doi [\apj] {10.3847/1538-4357/ac67a3}, \href {https://ui.adsabs.harvard.edu/abs/2022ApJ...930..144T} {930, 144}

\bibitem[\protect\citeauthoryear{{Tschudi} \& {Schmid}}{{Tschudi} \& {Schmid}}{2021}]{Tschudi2021A&A...655A..37T}
{Tschudi} C.,  {Schmid} H.~M.,  2021, \mn@doi [\aap] {10.1051/0004-6361/202141028}, \href {https://ui.adsabs.harvard.edu/abs/2021A&A...655A..37T} {655, A37}

\bibitem[\protect\citeauthoryear{{Verhamme}, {Kluska}, {Ferreira}, {Bollen}, {De Prins}, {Kamath}  \& {Van Winckel}}{{Verhamme} et~al.}{2024}]{Verhamme2024A&A...684A..79V}
{Verhamme} O.,  {Kluska} J.,  {Ferreira} J.,  {Bollen} D.,  {De Prins} T.,  {Kamath} D.,   {Van Winckel} H.,  2024, \mn@doi [\aap] {10.1051/0004-6361/202347708}, \href {https://ui.adsabs.harvard.edu/abs/2024A&A...684A..79V} {684, A79}

\bibitem[\protect\citeauthoryear{{Versteeg}, {Magalh{\~a}es}, {Haverkorn}, {Angarita}, {Rodrigues}, {Santos-Lima}  \& {Kawabata}}{{Versteeg} et~al.}{2023}]{Versteeg2023AJ....165...87V}
{Versteeg} M.~J.~F.,  {Magalh{\~a}es} A.~M.,  {Haverkorn} M.,  {Angarita} Y.,  {Rodrigues} C.~V.,  {Santos-Lima} R.,   {Kawabata} K.~S.,  2023, \mn@doi [\aj] {10.3847/1538-3881/aca8fd}, \href {https://ui.adsabs.harvard.edu/abs/2023AJ....165...87V} {165, 87}

\bibitem[\protect\citeauthoryear{{V{\"o}lschow}, {Banerjee}  \& {Hessman}}{{V{\"o}lschow} et~al.}{2014}]{Volschow2014A&A...562A..19V}
{V{\"o}lschow} M.,  {Banerjee} R.,   {Hessman} F.~V.,  2014, \mn@doi [\aap] {10.1051/0004-6361/201322111}, \href {https://ui.adsabs.harvard.edu/abs/2014A&A...562A..19V} {562, A19}

\bibitem[\protect\citeauthoryear{{Young}, {Alexander}, {Rosotti}  \& {Pinte}}{{Young} et~al.}{2022}]{Young2022MNRAS.513..487Y}
{Young} A.~K.,  {Alexander} R.,  {Rosotti} G.,   {Pinte} C.,  2022, \mn@doi [\mnras] {10.1093/mnras/stac840}, \href {https://ui.adsabs.harvard.edu/abs/2022MNRAS.513..487Y} {513, 487}

\bibitem[\protect\citeauthoryear{{Young}, {Stevenson}, {Nixon}  \& {Rice}}{{Young} et~al.}{2023}]{Young2023MNRAS.525.2616Y}
{Young} A.~K.,  {Stevenson} S.,  {Nixon} C.~J.,   {Rice} K.,  2023, \mn@doi [\mnras] {10.1093/mnras/stad2451}, \href {https://ui.adsabs.harvard.edu/abs/2023MNRAS.525.2616Y} {525, 2616}

\bibitem[\protect\citeauthoryear{{Zorotovic} \& {Schreiber}}{{Zorotovic} \& {Schreiber}}{2013}]{Zorotovic2013A&A...549A..95Z}
{Zorotovic} M.,  {Schreiber} M.~R.,  2013, \mn@doi [\aap] {10.1051/0004-6361/201220321}, \href {https://ui.adsabs.harvard.edu/abs/2013A&A...549A..95Z} {549, A95}

\bibitem[\protect\citeauthoryear{{de Boer} et~al.,}{{de Boer} et~al.}{2020}]{deBoer2020}
{de Boer} J.,  et~al., 2020, \mn@doi [\aap] {10.1051/0004-6361/201834989}, \href {https://ui.adsabs.harvard.edu/abs/2020A&A...633A..63D} {633, A63}

\bibitem[\protect\citeauthoryear{{van Boekel} et~al.,}{{van Boekel} et~al.}{2017}]{vanBoekel2017ApJ...837..132V}
{van Boekel} R.,  et~al., 2017, \mn@doi [\apj] {10.3847/1538-4357/aa5d68}, \href {https://ui.adsabs.harvard.edu/abs/2017ApJ...837..132V} {837, 132}

\bibitem[\protect\citeauthoryear{{van Holstein} et~al.,}{{van Holstein} et~al.}{2020}]{Holstein2020A&A...633A..64V}
{van Holstein} R.~G.,  et~al., 2020, \mn@doi [\aap] {10.1051/0004-6361/201834996}, \href {https://ui.adsabs.harvard.edu/abs/2020A&A...633A..64V} {633, A64}

\bibitem[\protect\citeauthoryear{{van Winckel}}{{van Winckel}}{2003}]{VanWinckel2003ARA&A..41..391V}
{van Winckel} H.,  2003, \mn@doi [\araa] {10.1146/annurev.astro.41.071601.170018}, \href {https://ui.adsabs.harvard.edu/abs/2003ARA&A..41..391V} {41, 391}

\makeatother
\end{thebibliography}



\appendix
\section{Presence of disc signal in the total intensity image}
\label{sec:scattered}

The total intensity frames of the IRAS\,08544-4431 ($I_{\rm tot}$) include both the stellar intensity and a minor contribution from the disc scattered light. \citep{Tschudi2021A&A...655A..37T} demonstrated the possibility of disentangling the disc intensity from the stellar signal by fitting the total intensity radial profile of the science target with a linear combination of the total intensity profile of a reference star and the polarized intensity profile of the disc. They noted that successful extraction of the disc intensity is feasible in cases where the disc exhibits a bright, narrow ring positioned sufficiently far from the star to ensure clear separation, yet remains within the strong speckle ring at $r \approx 100$ pixels \citep[control radius of the AO system;][]{Schmid2018A&A...619A...9S}.

\begin{figure*}
    
    \includegraphics[width=0.95\columnwidth]{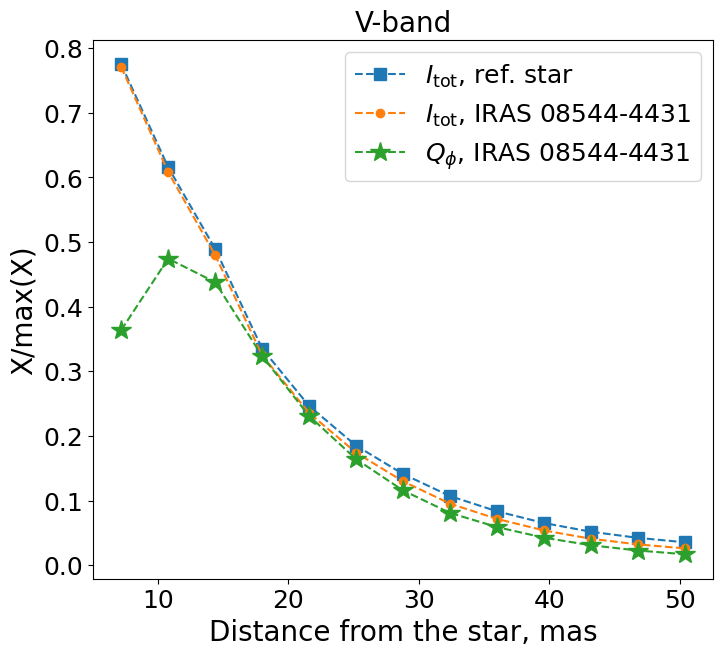}
    \includegraphics[width=0.95\columnwidth]{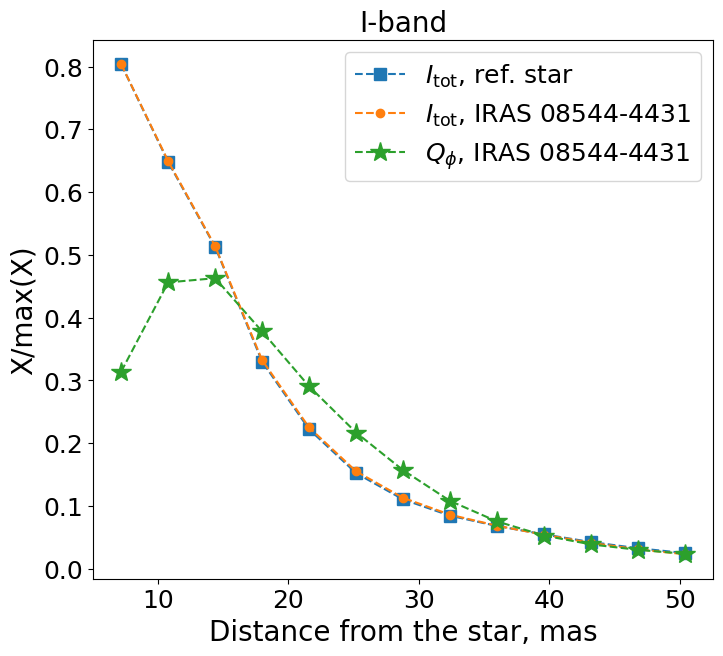}    
    
    \caption{Radial profiles of normalized total intensity ($I_{\rm tot}/max(I_{\rm tot})$) for IRAS\,08544-4431 (orange circles) and reference star HD\,83878 (blue squares), and normalised polarized intensity ($Q_\phi/max(Q_\phi)$) for IRAS\,08544-4431 (green stars). See Appendix~\ref{sec:scattered} for details.\label{fig:scattered}}
    
\end{figure*}

To investigate whether a similar disentanglement is achievable for our post-AGB binary system, we computed the radial brightness profiles for the total intensity ($I_{\rm tot}$) of IRAS\,08544-4431 and reference single star HD\,83878 (see Section~\ref{sec:profiles} for methodology details). The resulting profiles were normalized to the maximum intensity of each target and are presented in Fig.~\ref{fig:scattered}. Additionally, we provide the radial polarized brightness profile ($Q_{\phi}$), as it accurately traces the location of the disc.

Our analysis reveals that the radial profile of total intensity for IRAS\,08544-4431 closely aligns with that of the reference star. Therefore, we conclude that the intensity image is strongly dominated by the variable PSF of the star, and the disc intensity is too weak to be detected.

\bsp	
\label{lastpage}
\end{document}